\documentstyle[12pt]{article}

\begin{document}

\title{
Generalized uncertainty relations: \\
Theory, examples, and Lorentz invariance\thanks{
Supported in part by the Office of Naval Research
(Grant No.~N00014-93-1-0116)}
}

\author{
Samuel L.~Braunstein,\thanks{
Supported by a Feinberg Fellowship at the Weizmann Institute of Science.
Present address: Department of Chemical Physics, Weizmann Institute of
Science, 76100 Rehovot, Israel}{}~~Carlton M.~Caves,
and G.~J.~Milburn\thanks{
Permanent address: Department of Physics, University of Queensland,
St.~Lucia 4072, Australia}\\
\\
Center for Advanced Studies, Department of Physics and Astronomy,\\
University of New Mexico, Albuquerque, New Mexico 87131-1156
}

\date{\today}
\maketitle

\begin{abstract}
The quantum-mechanical framework in which observables are associated
with Hermitian operators is too narrow to discuss measurements of such
important physical quantities as elapsed time or harmonic-oscillator
phase.  We introduce a broader framework that allows us to derive
quantum-mechanical limits on the precision to which a parameter---e.g.,
elapsed time---may be determined via arbitrary data analysis of arbitrary
measurements on $N$ identically prepared quantum systems.  The limits
are expressed as generalized Mandelstam-Tamm uncertainty relations,
which involve the operator that generates displacements of the
parameter---e.g., the Hamiltonian operator in the case of elapsed
time.  This approach avoids entirely the problem of associating a
Hermitian operator with the parameter.  We illustrate the general
formalism, first, with nonrelativistic uncertainty relations for
spatial displacement and momentum, harmonic-oscillator phase and
number of quanta, and time and energy and, second, with Lorentz-invariant
uncertainty relations involving the displacement and Lorentz-rotation
parameters of the Poincar\'e group.
\end{abstract}

\newpage

\section{Introduction}
\label{intro}

The goal of quantitative experiments in physics is to determine a set
of parameters to some level of confidence. In general this determination
entails complex methods of data analysis applied to observed data.  From
this point of view, the conventional description of measurements in
quantum theory, tied to the use of Hermitian operators to represent
observable quantities, provides too narrow a framework, because for
many experimental parameters, time being an example, there is no suitable
Hermitian operator.

In this paper we employ a broader framework for describing the
quantum-me\-chan\-i\-cal determination of parameters such as time
\cite{BraunsteinCaves}.  In this framework measurements are described
in the most general way permitted by quantum mechanics---in terms of
so-called ``positive-operator-valued measures'' (POVMs).  The role of a
quantum measurement is to provide data from which one infers the
parameter of interest by classical methods of parameter estimation.
There is no need to associate a Hermitian operator with the parameter,
and generally there is no such Hermitian operator.  We derive quantum
restrictions on determining a parameter by considering optimal
measurements and optimal methods of parameter estimation.  The
quantum restrictions are stated as uncertainty relations that involve
the parameter and the operator that generates displacements of the
parameter, time and the Hamiltonian operator being an example.

Uncertainty relations are central to the interpretation of quantum
theory, yet in many cases of interest it is impossible to formulate
an uncertainty relation if one insists that both quantities have an
associated Hermitian operator.  Hilgevoord and Uffink \cite{Hilgevoord88}
give an excellent summary of the defects of standard uncertainty
relations and of the motivation for parameter-based uncertainty
relations.  Mandelstam and Tamm \cite{MandelstamTamm} derived the first
parameter-based uncertainty relation, for time and energy, by treating
elapsed time as a parameter to be determined by measurement of
a conventional observable that varies with time.  Helstrom \cite{Helstrom}
and Holevo \cite{Holevo} pioneered the modern study of parameter-based
uncertainty relations, by considering quantum restrictions on how
well one can determine a parameter from the results of general quantum
measurements described by POVMs.  Other authors
\cite{BraunsteinCaves,Hilgevoord88,Hilgevoord90,Dembo,BCNYAS} have
formulated parameter-based uncertainty relations in various contexts.

Here we present a general theory of parameter-based uncertainty
relations and explore in some detail the question of finding optimal
quantum measurements that achieve the lower bound set by the uncertainty
relation.  We devote Section~\ref{genup} to summarizing the framework
for quantum parameter estimation and the corresponding generalized
parameter-based uncertainty relations.  Section~\ref{genupmixed}
develops the general theory for mixed quantum states (density operators).
Section~\ref{genuppure} specializes the general theory to pure states
that are generated by a single-parameter unitary operator, a case
that occupies the remainder of the paper.  In Section~\ref{optmeas}
we develop a general description of global optimal measurements that
saturate the lower bound in the generalized uncertainty relation.
Section~\ref{exgenup} illustrates the parameter-based uncertainty
relations with various examples of nonrelativistic uncertainty relations:
spatial displacement and momentum in Section~\ref{xp}, harmonic-oscillator
phase and number of quanta in Section~\ref{phin}, and time and energy in
Section~\ref{tE}.  Section~\ref{lorentzup} applies the parameter-based
uncertainty relations to the displacement and Lorentz-rotation parameters
of the Poincar\'e group, leading ultimately to relativistically invariant
uncertainty relations for the invariant space-time interval of special
relativity and the boost and spatial-rotation parameters of Lorentz
transformations.  Section~\ref{conclusion} concludes with a brief
discussion.

\section{Generalized Uncertainty Relations}
\label{genup}

\subsection{Uncertainty relations for mixed states}
\label{genupmixed}

Consider $N$ replicas of a quantum system.  Each replica is prepared
in the same quantum state (density operator) $\hat\rho(X)$, which is
parametrized by the single parameter $X$.  In the following a
subscript $X$ on an expectation value denotes an expectation value
with respect to $\hat\rho(X)$.  Braunstein and Caves
\cite{BraunsteinCaves} consider a general smooth path on the space
of density operators,
\begin{equation}
\hat{\rho}(X)=\sum_jp_j|j\rangle\langle j| \;,
\label{rhodiagonal}
\end{equation}
where both the eigenvalues $p_j$ and the eigenvectors $|j\rangle$ can
change along the path.  A path is specified by giving the tangent vector
\begin{equation}
{d\hat\rho\over dX}=\sum_j{dp_j\over dX}|j\rangle\langle j|
-i[\hat h,\hat\rho]\equiv\hat\rho'\;.
\label{drho}
\end{equation}
The Hermitian operator $\hat h$, which can depend on $X$, generates
the infinitesimal changes in the eigenvectors of $\hat\rho(X)$:
\begin{equation}
\hat{\rho}(X+dX)=\hat\rho(X)+\hat\rho'(X)dX
=\sum_j(p_j+dp_j)e^{-idX\hat h}|j\rangle\langle j|e^{idX\hat h}\;.
\label{drhodiagonal}
\end{equation}
Notice that $\hat h$ can be replaced by
\begin{equation}
\Delta\hat h\equiv\hat h-\langle\hat h\rangle_X
\label{Deltah}
\end{equation}
in Eqs.~(\ref{drho}) and (\ref{drhodiagonal}) without changing the
path.

The most general measurement permitted by quantum mechanics
\cite{Holevo,Kraus,Peres1993} can be described by a set of bounded,
non-negative, Hermitian operators $\hat E(\xi)d\xi$ (generalizations
of projection operators), which are complete in the sense that
\begin{equation}
\int d\xi\,\hat E(\xi)=\hat1=\mbox{(unit operator)}\;.
\label{completeness}
\end{equation}
The quantity $\xi$ labels the ``results'' of the measurement; written
here as a single continuous real variable, it can be discrete or
multivariate.   The operators $\hat E(\xi)\,d\xi$ make up what is called
a ``positive-operator-valued measure'' (POVM).  The probability
distribution for result $\xi$, given the parameter $X$, is
\begin{equation}
p(\xi|X)= {\rm tr}\Bigl(\hat E(\xi)\hat\rho(X)\Bigr) \;.
\end{equation}
The properties of the POVM are just those needed to make $p(\xi|X)$
a normalized probability distribution.

Let $\xi_1,\ldots,\xi_N$ denote the results of measurements on the
$N$ replicas of our quantum system.  A general form of data analysis
uses a function
\begin{equation}
X_{\rm est} = X_{\rm est}(\xi_1,\ldots,\xi_N)
\label{estimator}
\end{equation}
to generate an estimate $X_{\rm est}$ for the parameter $X$, based on
the data $\xi_1,\ldots,\xi_N$ obtained from the measurements and
nothing else.

To characterize how precisely the $N$ measurements are able to determine
the parameter $X$, we need something a bit more complicated than the
obvious choice, the variance of the estimator,
$\langle(\Delta X_{\rm est})^2\rangle_X=
\langle(X_{\rm est}-\langle X_{\rm est}\rangle_X)^2\rangle_X$.  The
reason is that the variance does not take into account two important
possibilities.  First, even if the estimator has a small variance,
it might be systematically biased away from the true parameter
value---i.e., $\langle X_{\rm est}\rangle_X$ might not equal $X$---and
thus give a poor estimate.  Second, the estimator might have different
``units'' from the parameter, thus making it difficult to interpret
the variance of the estimator as a measure of precision in determining
$X$.  Both the amount of bias and the difference in units can depend
on the parameter, i.e., on location along the path.  To remedy these
difficulties, we quantify the estimate's deviation from the parameter
by \cite{BraunsteinCaves}
\begin{equation}
\delta X\equiv{X_{\rm est}\over|d\langle X_{\rm est}\rangle_X/dX|}-X \;.
\label{deltaX}
\end{equation}
The derivative $d\langle X_{\rm est}\rangle_X/dX$ removes the local
difference in the ``units'' of the estimator and the parameter, and
then the units-corrected estimator is compared to the parameter $X$,
not to the mean value of the estimator.  As a statistical measure of
the precision of the estimation, we use the second moment of $\delta X$.

There is a lower bound on the second moment of $\delta X$:
\begin{equation}
\langle(\delta X)^2\rangle_X\ge{1\over NF(X)}\ge{1\over N(ds/dX)^2}\;.
\label{BCbound}
\end{equation}
Braunstein and Caves \cite{BraunsteinCaves} derive the ultimate lower
bound in two steps, in contrast to derivations by Helstrom \cite{Helstrom}
and Holevo [5(Chap.~VI.2)], both of whom proceed to the ultimate lower
bound in a single step that obscures the conditions for achieving
the ultimate lower bound.  The two steps in the Braunstein-Caves
derivation are displayed as the two inequalities in Eq.~(\ref{BCbound}).
The first inequality is a bound that applies to all estimators
$X_{\rm est}$ for a fixed probability distribution $p(\xi|X)$, i.e.,
for a fixed quantum measurement.  The second inequality is a bound that
applies to all quantum measurements.

In the first inequality in Eq.~(\ref{BCbound}),
\begin{equation}
F(X)\equiv\int d\xi\,{1\over p(\xi|X)}
\left({\partial p(\xi|X)\over\partial X}\right)^2
\label{Fisherinfo}
\end{equation}
is the {\it Fisher information\/} \cite{CramerRao} associated with the
probability distribution $p(\xi|X)$.  The first inequality is an
expression of the Cram\'er-Rao bound of classical estimation theory
\cite{CramerRao}, which places a lower bound on the variance of {\it any\/}
estimator $X_{\rm est}$ that is applied to data drawn from the
distribution $p(\xi|X)$.  An estimator that saturates the first
inequality in Eq.~(\ref{BCbound})---and, hence, attains the Cram\'er-Rao
bound---is called an {\it efficient\/} estimator.  The lower bound in
the first inequality can always be achieved asymptotically for large
$N$ by using maximum-likelihood estimation \cite{Fisher}, but except
for special distributions, there is {\it no\/} efficient estimator for
finite values of $N$.

The second inequality in Eq.~(\ref{BCbound}) holds for any POVM
$\hat E(\xi)\,d\xi$.  The second inequality is written in terms of a line
element $ds^2$, which defines a ``statistical distance'' \cite{Wootters}
that measures the distinguishability of neighboring quantum states and
provides a natural Riemannian geometry on the space of density operators.
The explicit form that Braunstein and Caves \cite{BraunsteinCaves} (see
also [4,5(Chap.~VI.2)]) find for the line element is
\begin{equation}
ds^2/dX^2=
\left\langle\Bigl({\cal L}_{\hat\rho}(\hat\rho')\Bigr)^2\right\rangle_X=
{\rm tr}\Bigl(\hat\rho'{\cal L}_{\hat\rho}(\hat\rho')\Bigr)\;,
\label{statdistance}
\end{equation}
where ${\cal L}_{\hat\rho}$ is a super-operator that, in the basis
that diagonalizes $\hat\rho$, takes the form
\begin{equation}
{\cal L}_{\hat\rho}(\hat O)=
\sum_{\{j,k|p_j+p_k\ne0\}}{2\over p_j+p_k}O_{jk}|j\rangle\langle k|\;.
\label{scL}
\end{equation}
If $\hat\rho$ has no zero eigenvalues, ${\cal L}_{\hat\rho}$ is
the inverse of the super-operator defined by
${\cal R}_{\hat\rho}(\hat O)\equiv{1\over2}(\hat\rho\hat O+\hat O\hat\rho)$;
hence, ${\cal L}_{\hat\rho}$ was denoted ${\cal R}_{\hat\rho}^{-1}$
by Braunstein and Caves \cite{BraunsteinCaves}.  Helstrom
\cite{Helstrom} and Holevo [5(Chap.~VI.2)] call
${\cal L}_{\hat\rho}(\hat\rho')$ the ``symmetric logarithmic
derivative'' of $\hat\rho$.  The lower bound in the second inequality
can be achieved by using a POVM such that the operators $\hat E(\xi)$
are one-dimensional projection operators onto orthonormal eigenstates
of the Hermitian operator ${\cal L}_{\hat\rho}(\hat\rho')$
\cite{BraunsteinCaves}.  The conditions given here and above for achieving
the two lower bounds in Eq.~(\ref{BCbound}) are sufficient, though they
are not always necessary.

The line element~(\ref{statdistance}) on the space of density operators
arises here from quantifying precisely the quantum restrictions on
determining a parameter---and, hence, the quantum restrictions on the
statistical distinguishability of neighboring density operators
$\rho(X)$ and $\rho(X+dX)$.  The same line element can also be gotten
by defining a natural metric on density operators in terms of the
correlation between pairs of conventional observables.  The reader
interested in this alternative route to the metric can find it spelled
out in \cite{BCNYAS,BCNottingham}, together with references to related
work.

Although not pointed out by Braunstein and Caves, the lower
bound~(\ref{BCbound}) does not improve if one allows measurements that
do not factor into separate measurements on each of the $N$ replicas.
One can see this by treating the $N$ replicas as a single composite
system with density operator
\begin{equation}
\hat{\rho}^{(N)}(X)=\hat{\rho}(X)\otimes \cdots \otimes \hat{\rho}(X) \;.
\label{tensorproduct}
\end{equation}
Applied to this composite system, the bound~(\ref{BCbound}) takes a
form $\langle(\delta X)^2\rangle_X\ge(dX/ds^{(N)})^2$ that holds for
all quantum measurements on the product space of the composite system.
It is not difficult to show, however, that for $N$-replica product states,
the line element on the product space reduces to $N$ times the
single-replica line element, i.e.,
\begin{equation}
{(ds^{(N)})^2\over dX^2}={\rm tr}\Biggl({d\hat\rho^{(N)}\over dX}\,
{\cal L}_{\hat\rho^{(N)}}\biggl({d\hat\rho^{(N)}\over dX}\biggr)\Biggr)=
N{\rm tr}\Bigl(\hat\rho'{\cal L}_{\hat\rho}(\hat\rho')\Bigr)\;,
\end{equation}
thus giving the same lower bound as in Eq.~(\ref{BCbound}).  This
result provides a limited answer to a question raised by Peres and
Wootters \cite{PeresWootters}: when a composite system is made up of
replicas all prepared in the same quantum state, can measurements on
the composite system better distinguish states than can separate
measurements on each of the replicas?  For the very special case of
two neighboring states, the answer is no.

We pause at this point to take stock of what has already been presented.
The bound~(\ref{BCbound}), together with Eq.~(\ref{statdistance}), is
a general species of uncertainty relation, which restricts one's ability
to determine a parameter from the results of quantum measurements.  This
uncertainty relation applies to mixed quantum states, allows for
measurements that are not described by projection operators, and includes
the possibility of multiple measurements.  On the other hand, precisely
because this uncertainty relation is so general, we find it instructive
in what follows to specialize in three ways, thus permitting us to make
closer contact with standard uncertainty relations.

For the first specialization we assume that the eigenvalues of the density
operator do not change along the path---i.e., $dp_j=0$ in
Eq.~(\ref{drhodiagonal})---which simplifies Eq.~(\ref{drho}) to
\begin{equation}
\hat\rho'=-i[\hat h,\hat\rho]=-i[\Delta\hat h,\hat\rho]\;.
\label{rhoprimetwo}
\end{equation}
This first specialization means that the path is generated by a
unitary transformation; keep in mind, however, that we still allow
the local generator of the transformation, $\hat h$, to depend on $X$.
As a consequence of this first specialization, we can write
\begin{equation}
{\cal L}_{\hat\rho}(\hat\rho')=
2i\sum_{\{j,k|p_j+p_k\ne0\}}{p_j-p_k\over p_j+p_k}
\Delta h_{jk}|j\rangle\langle k|\equiv2\widehat{\delta h}\;,
\label{deltah}
\end{equation}
where we introduce $\widehat{\delta h}$ as a shorthand for
${1\over2}{\cal L}_{\hat\rho}(\hat\rho')$, and the line
element~(\ref{statdistance}) becomes
\begin{equation}
{ds^2\over dX^2}=4\langle(\widehat{\delta h})^2\rangle_X=
2\sum_{j,k}(p_j+p_k)\!\left({p_j-p_k\over p_j+p_k}\right)^{\!2}\!
|\Delta h_{jk}|^2\le4\langle(\Delta\hat h)^2\rangle_X\;.
\label{statdistancetwo}
\end{equation}
Notice that in this line element we can drop the restriction on
the sum, since under any procedure for approaching the boundary
on which one or more eigenvalues of $\hat\rho$ vanishes, the terms
for which $p_j+p_k=0$ do not contribute.

A consequence of the last inequality in Eq.~(\ref{statdistancetwo}) is a
parameter-based uncertainty relation [1,4,5(Chaps.~III.2, IV.7, and VI.3)],
\begin{equation}
\langle(\delta X)^2\rangle_X\langle(\Delta\hat h)^2\rangle_X\ge
{1\over 4N}\;,
\label{pureup1}
\end{equation}
which, since it involves the variance of $\hat h$, resembles standard
uncertainty relations, except that $X$ is a parameter and the relation
holds for multiple measurements.  The corresponding uncertainty relation
involving $\widehat{\delta h}$,
\begin{equation}
\langle(\delta X)^2\rangle_X\langle(\widehat{\delta h})^2\rangle_X\ge
{1\over 4N}\;,
\label{deltahup}
\end{equation}
is stricter [1,5(Chap.~VI.3)], unless equality holds in
Eq.~(\ref{statdistancetwo}).  Equality is equivalent to the condition
that $p_jp_k|\Delta h_{jk}|^2=0$ for all $j$ and $k$.  In particular,
equality holds if $\hat\rho$ is a pure state, but never holds if
$\hat\rho$ has no zero eigenvalues (except in the trivial case
$\Delta\hat h=0$).

\subsection{Uncertainty relations for pure states}
\label{genuppure}

The second specialization is to assume that
$\hat\rho(X)=|\psi_X\rangle\langle\psi_X|$ is a pure state.  This
assumption implies the first one, which is incorporated in
Eq.~(\ref{rhoprimetwo}), since a path on the pure states must be
generated by a unitary transformation.  Normalization implies that
\begin{equation}
0={d\over dX}\langle\psi_X|\psi_X\rangle=
\langle\psi_X|\!\left({d|\psi_X\rangle\over dX}\right)+
\left({d\langle\psi_X|\over dX}\right)\!|\psi_X\rangle=
2{\rm Re}\Biggl(\langle\psi_X|\biggl({d|\psi_X\rangle\over dX}\biggr)\Biggr)\;,
\label{normcond}
\end{equation}
but the freedom to multiply $|\psi_X\rangle$ by a phase factor means
that ${\rm Im}[\langle\psi_X|(d|\psi_X\rangle/dX)]$ can be chosen
arbitrarily.  Using Eqs.~(\ref{rhoprimetwo}) and (\ref{normcond}),
one can show that
\begin{equation}
\left({d|\psi_X\rangle\over dX}\right)_{\!\!\perp}\equiv
{d|\psi_X\rangle\over dX}-
|\psi_X\rangle\langle\psi_X|\!\left({d|\psi_X\rangle\over dX}\right)=
-i\Delta\hat h|\psi_X\rangle\;,
\label{dpsiperp}
\end{equation}
where $(d|\psi_X\rangle/dX)_{\!\perp}$ is the projection of
$d|\psi_X\rangle/dX$ orthogonal to $|\psi_X\rangle$.
Equation~(\ref{rhoprimetwo}) can now be written as
\begin{equation}
\hat\rho'=-i[\Delta\hat h,\hat\rho]=
\left({d|\psi_X\rangle\over dX}\right)_{\!\!\perp}\!\!\langle\psi_X|+
|\psi_X\rangle\!\left({d\langle\psi_X|\over dX}\right)_{\!\!\perp}\;.
\end{equation}
A convenient phase choice,
\begin{equation}
\langle\psi_X|\!\left({d|\psi_X\rangle\over dX}\right)=
-i\langle\hat h\rangle_X\;,
\end{equation}
leads to a Schr\"odinger-like equation for $|\psi_X\rangle$:
\begin{equation}
{d|\psi_X\rangle\over dX}=-i\hat h|\psi_X\rangle\;.
\label{dpsi}
\end{equation}
Notice that the phase freedom in $|\psi_X\rangle$ is equivalent to
the freedom to add a multiple of the unit operator to $\hat h$.

Applying our second assumption to Eq.~(\ref{deltah}), one finds that
\begin{equation}
\widehat{\delta h}={1\over2}{\cal L}_{\hat\rho}(\hat\rho')=
-i[\Delta\hat h,\hat\rho]=
\left({d|\psi_X\rangle\over dX}\right)_{\!\!\perp}\!\!\langle\psi_X|+
|\psi_X\rangle\!\left({d\langle\psi_X|\over dX}\right)_{\!\!\perp}=\hat\rho'\;.
\label{deltahtwo}
\end{equation}
Thus, for pure states, the line element~(\ref{statdistancetwo}) for
statistical distance reduces to
\begin{equation}
{ds^2\over dX^2}=
4\left({d\langle\psi_X|\over dX}\right)_{\!\!\perp}
\!\left({d|\psi_X\rangle\over dX}\right)_{\!\!\perp}=
4\langle(\Delta\hat h)^2\rangle_X\;,
\label{s2variance}
\end{equation}
which implies, as indicated above, that we can restrict our attention
to the uncertainty relation~(\ref{pureup1}).

One expects statistical distance, which measures the distinguishability
of states, to be related to the inner product and thus to the
Hilbert-space angle between pure states.  The square of the
infinitesimal Hilbert-space angle $d\theta$ between neighboring
states $|\psi_X\rangle$ and $|\psi_{X+dX}\rangle$ is
\begin{equation}
d\theta^2=[\cos^{-1}(|\langle\psi_X|\psi_{X+dX}\rangle|)]^2=
1-|\langle\psi_X|\psi_{X+dX}\rangle|^2\;.
\label{Hangle}
\end{equation}
The line element $d\theta^2$ defines a natural Riemannian metric,
called the Fubini-Study metric \cite{AnandanAharonov,Anandan,Gibbons},
on the manifold of Hilbert-space rays.  Using Eq.~(\ref{normcond})
and the further consequence of normalization,
\begin{equation}
0={1\over2}{d^2\over dX^2}\langle\psi_X|\psi_X\rangle=
\left({d\langle\psi_X|\over dX}\right)\!
\left({d|\psi_X\rangle\over dX}\right)+
{\rm Re}\Biggl(\langle\psi_X|
\biggl({d^2|\psi_X\rangle\over dX^2}\biggr)\Biggr)\;,
\end{equation}
one finds that
\begin{eqnarray}
d\theta^2&=&dX^2\!\left[
\left({d\langle\psi_X|\over dX}\right)\!
\left({d|\psi_X\rangle\over dX}\right)-
\left|\langle\psi_X|\!
\left({d|\psi_X\rangle\over dX}\right)\right|^2\right]\nonumber\\
&=&dX^2\!\left({d\langle\psi_X|\over dX}\right)_{\!\!\perp}
\!\left({d|\psi_X\rangle\over dX}\right)_{\!\!\perp}={1\over4}ds^2\;,
\label{Hangles2}
\end{eqnarray}
which means that Hilbert-space angle is half the statistical distance
defined here.

The third and final specialization is to assume that the generator
$\hat h$ is independent of $X$.  This assumption allows us to
integrate Eq.~(\ref{dpsi}) and to write the path on Hilbert space as
being generated by a single-parameter unitary operator,
\begin{equation}
|\psi_X\rangle=e^{-iX\hat h}|\psi_0\rangle\;,
\label{psiX}
\end{equation}
where $|\psi_0\rangle$ is a fiducial state at $X=0$.  Moreover,
this assumption guarantees that the expectation value of any function
of $\hat h$ is independent of $X$; thus we can drop the subscript $X$
from the mean and variance of $\hat h$.  In particular, we can write
the uncertainty relation~(\ref{pureup1}) as
\begin{equation}
\langle(\delta X)^2\rangle_X\langle(\Delta\hat h)^2\rangle\ge{1\over 4N}\;.
\label{pureup}
\end{equation}
It is this parameter-based uncertainty relation that occupies us for
the remainder of this paper.  As noted above, this uncertainty relation
resembles the standard uncertainty relation, except that the relation
holds for multiple measurements and $X$ is a parameter, not necessarily
corresponding to any Hermitian operator.

\section{Global Optimal Measurements}
\label{optmeas}

\subsection{General considerations}
\label{optmeasA}

The chain of inequalities leading to the uncertainty
relation~(\ref{pureup}) consists of the two inequalities leading
to the statistical distance in Eq.~(\ref{BCbound}) and the inequality
involving the generator $\hat h$ in Eq.~(\ref{statdistancetwo}).
The first inequality in Eq.~(\ref{BCbound}) can be saturated
asymptotically for large $N$ by use of maximum-likelihood estimation,
and the inequality in Eq.~(\ref{statdistancetwo}) is saturated for pure
states.  Thus the question of achieving equality in the uncertainty
relation~(\ref{pureup}), provided one allows for many measurements $N$,
reduces to finding an {\it optimal measurement}, i.e., one that
saturates the second inequality in Eq.~(\ref{BCbound}).  Notice that
since the variance of $\hat h$ is constant as a consequence of our
third assumption, optimal measurements lead to a maximum Fisher
information that is constant along the path.

As indicated above, one such optimal measurement uses a POVM such that the
operators $\hat E(\xi)$ are one-dimensional projection operators onto
orthonormal eigenstates of the Hermitian operator
${\cal L}_{\hat\rho}(\hat\rho')=2\widehat{\delta h}$
[cf.~Eq.~(\ref{deltahtwo})].  This measurement has the defect, however,
that it generally depends on $X$, thus requiring one to know the value of
the parameter one is trying to estimate before choosing the optimal
measurement.  Our goal here is to find a {\it global\/} measurement,
independent of $X$, that is optimal all along the path.  We seek such
a global optimal measurement in terms of a POVM $\hat E(x)\,dx$, where
the measurement results are labeled by a single real number $x$ that has
the same range of values as $X$.  As we discuss further in
Section~\ref{optmeasB}, we can hope to find an optimal measurement of
this form only if the generator $\hat h$ is non-degenerate; if the
spectrum of $\hat h$ has degeneracies, an optimal measurement must
acquire information beyond that which can be described by a single
real number.

The POVM $\hat E(x)\,dx$ must, of course, be complete, which means
that
\begin{equation}
\hat1=\int dx\,\hat E(x)\;.
\label{Extwo}
\end{equation}
The probability density for result $x$, given the parameter $X$, is
\begin{equation}
p(x|X)=\langle\psi_X|\hat E(x)|\psi_X\rangle=
\langle\psi_0|e^{iX\hat h}\hat E(x)e^{-iX\hat h}|\psi_0\rangle\;.
\label{pxX}
\end{equation}
As noted above, global optimal measurements lead to a Fisher
information~(\ref{Fisherinfo}) that is independent of $X$.  This
suggests that we require that $p(x|X)$ be a function only of $x-X$,
which means the POVM must satisfy a ``displacement'' property
\begin{equation}
e^{iX\hat h}\hat E(x)e^{-iX\hat h}=\hat E(x-X)\;.
\label{Exthree}
\end{equation}
Measurements that satisfy properties~(\ref{Extwo}) and (\ref{Exthree})
are called {\it covariant\/} by Holevo \cite{Holevo}.

We restrict our search for global optimal measurements to POVMs
that have one additional property: the POVM consists of multiples
of ``projection operators'' onto (generally unnormalizable) states
$|x\rangle$,
\begin{equation}
\hat E(x)\,dx={dx\over C}\,|x\rangle\langle x|
\label{Exone}
\end{equation}
($C$ is a real constant).  The motivation for this assumption is that
measurements not described by one-dimensional ``projectors'' have less
resolution \cite{Hall94}, but it would be useful to make this motivation
precise or to investigate whether covariant measurements that do not
satisfy property~(\ref{Exone}) can be optimal.  Notice that we do {\it not\/}
require that the states $|x\rangle$ be orthogonal, and if they are not,
they are necessarily {\it over\/}complete.  The constant $C$ could be
absorbed into the states $|x\rangle$, but it is useful to leave it
free so that these states can be given conventional normalizations in
the examples of Section~\ref{exgenup}.

Without loss of generality we can discard the freedom to re-phase
the states $|x\rangle$, because the POVM is unaffected by re-phasing,
and thus replace the displacement property of the POVM with a displacement
requirement on the states,
\begin{equation}
e^{-iX\hat h}|x\rangle=|x+X\rangle\;.
\label{Exfour}
\end{equation}
This displacement property, written as
\begin{equation}
\langle x|e^{-iX\hat h}|\psi\rangle=\langle x-X|\psi\rangle=
e^{-X\partial/\partial x}\langle x|\psi\rangle\;,
\end{equation}
is equivalent to saying that in the $x$ representation, $\hat h$ is
represented by a derivative:
\begin{equation}
\hat h\Longleftrightarrow{1\over i}{\partial\over\partial x}\;.
\end{equation}
The probability density~(\ref{pxX}) can be written as
\begin{equation}
p(x|X)=|\psi_X(x)|^2=|\psi_0(x-X)|^2\equiv p(x-X)\;,
\end{equation}
where
\begin{equation}
\psi_X(x)\equiv{1\over\sqrt C}\langle x|\psi_X\rangle=
{1\over\sqrt C}\langle x-X|\psi_0\rangle=\psi_0(x-X)
\end{equation}
is the ``wave function'' of the state vector $|\psi_X\rangle$ in
the $x$ representation.

Equations~(\ref{Extwo}), (\ref{Exone}), and (\ref{Exfour}) are the three
properties that we require of the POVM $\hat E(x)\,dx$.  Holevo
[5(Chap.~IV.7)] considers the same sorts of measurements; his treatment,
while more rigorous mathematically than ours, is inaccessible to many
physicists.  The three properties are preserved by a ``gauge
transformation,'' which replaces the states $|x\rangle$ with states
\begin{equation}
e^{if(\hat h)}|x\rangle\;,
\label{gaugetrans}
\end{equation}
where $f$ is an arbitrary real-valued function.  This gauge freedom
plays an important role, as we discuss further in Section~\ref{optmeasB}
and in the examples of Section~\ref{exgenup}.

If the POVM $\hat E(x)\,dx$ is an optimal measurement, then it
saturates the second inequality in Eq.~(\ref{BCbound}), which
simplifies to
\begin{equation}
\int dx\,{[\,p'(x)]^2\over p(x)}=F\le
{ds^2\over dX^2}=4\langle(\Delta\hat h)^2\rangle\;.
\label{BCboundtwo}
\end{equation}
In this inequality we put the Fisher information $F$ in a new form, which
applies to a covariant measurement and which is explicitly independent
of $X$.

For a measurement described by the one-dimensional ``projectors''
$|x\rangle\langle x|$, the necessary and sufficient condition for an
optimal measurement, as shown in \cite{BraunsteinCaves}, is that
\begin{equation}
{\rm Im}\Biggl(\langle\psi_X|x\rangle\langle x|
\biggl({d|\psi_X\rangle\over dX}\biggr)_{\!\perp}\,\Biggr)=0\;\;\;
\mbox{for all $x$ and all $X$.}
\label{condeq}
\end{equation}
Using Eqs.~(\ref{dpsiperp}) and (\ref{Exthree}) and writing
\begin{equation}
{1\over\sqrt C}\langle x|\psi_0\rangle=\psi_0(x)=r(x)e^{i\Theta(x)}\;,
\label{psizero}
\end{equation}
\begin{equation}
r(x)=|\psi_0(x)|=\sqrt{p(x)}\;,
\end{equation}
one can recast condition~(\ref{condeq}) as
\begin{equation}
0={1\over C}
{\rm Im}(i\langle\psi_0|x\rangle\langle x|\Delta\hat h|\psi_0\rangle)=
r^2(x)[\Theta'(x)-\langle\hat h\rangle]\;\;\;\mbox{for all $x$,}
\label{condeqtwo}
\end{equation}
which is equivalent to $\Theta(x)=\langle\hat h\rangle x+\mbox{constant}$.
After discarding the irrelevant overall phase due to the constant, the
resulting wave function is
\begin{equation}
\psi_0(x)=r(x)e^{i\langle\hat h\rangle x}\;.
\label{optwf}
\end{equation}
The POVM $\hat E(x)\,dx$ thus describes a global optimal measurement
if and only if the wave function $\psi_0(x)$ of the fiducial state is
(up to an overall phase) an arbitrary real function times a simple
phase factor that accounts for the expectation value of $\hat h$.
For a fiducial state whose wave function has a phase that is nonlinear
in $x$ for all choices of the states $|x\rangle$, we cannot rule out
the existence of a global optimal measurement, but we can say that
any measurement that satisfies properties~(\ref{Extwo}), (\ref{Exone}),
and (\ref{Exfour}) is not optimal.

We can get at condition~(\ref{condeqtwo}) directly by calculating the
mean and variance of $\hat h$ in the $x$ representation, again writing
$\psi_0(x)$ as
in Eq.~(\ref{psizero}):
\begin{equation}
\langle\hat h\rangle=\int dx\,\psi_0^\ast(x)
{1\over i}{\partial\over\partial x}\psi_0(x)=
\int dx\,p(x)\Theta'(x)\;,
\label{meanh}
\end{equation}
\begin{equation}
\langle(\Delta\hat h)^2\rangle=
\int dx\,\left|
\left({\partial\over\partial x}-i\langle\hat h\rangle\!\right)
\!\psi_0(x)\right|^2=
{1\over 4}\int dx\,{[\,p'(x)]^2\over p(x)}+
\int dx\,p(x)[\Theta'(x)-\langle\hat h\rangle]^2\;.
\label{varianceh}
\end{equation}
This expression for $\langle(\Delta\hat h)^2\rangle$ connects the
the Cram\'er-Rao bound of classical estimation theory [first inequality
in Eq.~(\ref{BCbound})] to the requirements of quantum theory
[second inequality in Eq.~(\ref{BCbound})].  A glance at
Eq.~(\ref{BCboundtwo}) reminds one that the first term in
$\langle(\Delta\hat h)^2\rangle$ is one-quarter of the Fisher
information; moreover, one recognizes that for an optimal measurement
this first term must attain its maximum value, which is the variance
of $\hat h$.  Thus, for an optimal measurement, the second term in
$\langle(\Delta\hat h)^2\rangle$, which is the variance of $\Theta'(x)$
with respect to $p(x)$, must be zero; vanishing of the second term is
precisely the condition~(\ref{condeqtwo}).

It is instructive to consider in some detail a special case of the
uncertainty relation~(\ref{pureup}), because in this special case one
finds the closest connection between our parameter-based uncertainty
relations and standard uncertainty relations.  Before considering this
special case, however, it is useful to note that the mean and variance
of the measurement result $x$ are given by
\begin{equation}
\langle x\rangle_X=\int dx\, xp(x-X)=X+\int dx\, xp(x)=X+\langle x\rangle_0\;,
\label{meanx}
\end{equation}
\begin{equation}
\langle(\Delta x)^2\rangle_X=
\int dx\,(x-\langle x\rangle_X)^2p(x-X)=
\int dx\,(x-\langle x\rangle_0)^2p(x)=\langle(\Delta x)^2\rangle\;.
\label{variancex}
\end{equation}
The mean value of $x$ with respect to the fiducial state,
$\langle x\rangle_0$, globally biases the mean $\langle x\rangle_X$
away from the parameter.  The variance of $x$ is independent of $X$.

To introduce our special case, suppose that one makes $N$ measurements
described by the POVM $\hat E(x)\,dx$ and that one estimates the
parameter $X$ as the sample mean of the data, with the global bias
removed, i.e.,
\begin{equation}
X_{\rm est}={1\over N}\sum_{i=1}^N(x_i-\langle x\rangle_0)\;.
\label{samplemean}
\end{equation}
This estimator is {\it unbiased}, i.e.,
$\langle X_{\rm est}\rangle_X=\langle x\rangle_X-\langle x\rangle_0=X$,
and thus the deviation~(\ref{deltaX}) becomes
$\delta X=X_{\rm est}-X=\Delta X_{\rm est}$.  In addition, the efficiency
of this estimator is independent of $N$, because the mean-square deviation
decreases as $1/N$:
\begin{equation}
\langle(\delta X)^2\rangle=\langle(\Delta X_{\rm est})^2\rangle=
\langle(\Delta x)^2\rangle\;.
\end{equation}
The resulting special case of the uncertainty relation~(\ref{pureup}) is
\begin{equation}
\langle(\Delta x)^2\rangle\langle(\Delta\hat h)^2\rangle=
N\langle(\delta X)^2\rangle\langle(\Delta\hat h)^2\rangle
\ge{1\over4}\;.
\label{pureup2}
\end{equation}
The uncertainty relation for this estimator is identical to a standard
uncertainty relation for the measurement result $x$, the only difference
being that the states $|x\rangle$ are generally not the eigenstates of
any Hermitian operator.

Equality in the uncertainty relation~(\ref{pureup2}) requires saturating
both inequalities in Eq.~(\ref{BCbound}).  Saturating the second
inequality---i.e., making $F=4\langle(\Delta\hat h)^2\rangle$ [see
Eq.~(\ref{BCboundtwo})]---means that the fiducial wave function
$\psi_0(x)$ has the form~(\ref{optwf}).  Saturating the first inequality
means that the sample mean~(\ref{samplemean}) is an efficient estimator
for all values of $N$ and, in particular, that $x-\langle x\rangle_0$,
the measurement result with the global bias removed, is itself an
efficient estimator for $N=1$.  We can determine the resulting conditions
by specializing the proof of the Cram\'er-Rao bound to the case of a
single measurement with $x-\langle x\rangle_0$ as the estimator.  We
first write the mean of the estimator in the form
\begin{equation}
\langle x\rangle_X-\langle x\rangle_0=
\int dx\,(x-\langle x\rangle_0)p(x-X)=X\;,
\end{equation}
Differentiating this expression with respect to $X$ and using
\begin{equation}
0={d\over dX}\int dx\,p(x-X)=-\int dx\,p'(x-X)
\end{equation}
leads to
\begin{eqnarray}
1&=&-\int dx\,(x-\langle x\rangle_0)p'(x-X)\nonumber\\
&=&-\int dx\,(x-\langle x\rangle_0-X)p'(x-X)\nonumber\\
&=&-\int dx\,(x-\langle x\rangle_0)p'(x)=
-\int dx\,p(x)(x-\langle x\rangle_0){d\ln p(x)\over dx}\;.
\end{eqnarray}
Squaring this expression and using the Schwarz inequality yields
\begin{eqnarray}
1&=&
\left(\int dx\,p(x)(x-\langle x\rangle_0){d\ln p(x)\over dx}\right)^{\!2}
\nonumber\\
&\le&\left(\int dx\,(x-\langle x\rangle_0)^2p(x)\right)\!
\left(\int dx\,p(x)\!\left({d\ln p(x)\over dx}\right)^{\!2}\right).
\label{CRproof}
\end{eqnarray}
Re-writing the expression for the Fisher information in
Eq.~(\ref{BCboundtwo}) as
\begin{equation}
F=\int dx\,{[\,p'(x)]^2\over p(x)}=
\int dx\,p(x)\!\left({d\ln p(x)\over dx}\right)^2
\label{Fisherinfotwo}
\end{equation}
shows that Eq.~(\ref{CRproof}) is the classical $N=1$ bound on the
estimator $x$:
\begin{equation}
\langle(\Delta x)^2\rangle=\int dx\,(x-\langle x\rangle_0)^2p(x)
\ge{1\over F}\;.
\end{equation}
The condition for saturating this bound, which comes from the Schwarz
inequality in Eq.~(\ref{CRproof}), is that
\begin{equation}
{d\ln p(x)\over dx}=-\lambda(x-\langle x\rangle_0)\;\;\;
\Longrightarrow\;\;\;
p(x)\propto e^{-\lambda(x-\langle x\rangle_0)^2/2}\;,
\end{equation}
where $\lambda$ is a constant.

The result of these considerations is that equality in the uncertainty
relation~(\ref{pureup2}) can be achieved if and only if the fiducial
wave function has the form~(\ref{optwf}), with $p(x)=r^2(x)$ being a
Gaussian.  These Gaussian states are analogous to the minimum-uncertainty
states that give equality in the standard uncertainty relation.  Thus
our formalism of parameter-based uncertainty relations contains within
itself, in the special case of the estimator being the sample mean,
the standard uncertainty relation and the associated minimum-uncertainty
states.  Two points deserve mention.  First, for most generators $\hat h$,
there are restrictions on the form of the wave function; these restrictions,
which are discussed in Section~\ref{optmeasB} and in the examples of
Section~\ref{exgenup}, generally prevent one from choosing a Gaussian for
$p(x)$ and thus mean that there are no states that yield equality in the
uncertainty relation~(\ref{pureup2}).  Second, the restriction to Gaussian
wave functions is a consequence of using the sample mean as the estimator.
If one allows other estimators, the conditions on the fiducial wave function
are weaker.  Specifically, as we have seen, in the limit of large $N$,
where maximum-likelihood estimation is asymptotically efficient, the
condition for saturating the uncertainty relation~(\ref{pureup}) is that
the fiducial wave function have the form~(\ref{optwf}).

\subsection{The $x$ representation}
\label{optmeasB}

Up till now, it has not been necessary to construct explicitly states
$|x\rangle$ that satisfy the completeness and displacement properties.
Such a construction depends on the eigenvalue spectrum of $\hat h$.
Suppose that we write the eigenvalue equation for $\hat h$ as
\begin{equation}
\hat h|h,\alpha\rangle=h|h,\alpha\rangle\;,
\end{equation}
where we allow for the possibility of degeneracies by including a
degeneracy label $\alpha$.  The orthonormal eigenstates $|h,\alpha\rangle$
satisfy a completeness relation
\begin{equation}
\hat1=\sum_{h,\alpha}|h,\alpha\rangle\langle h,\alpha|\;.
\end{equation}
The displacement property~(\ref{Exfour}), with $x=0$ and $X=x$, becomes
\begin{equation}
\langle h,\alpha|x\rangle=e^{-ixh}\langle h,\alpha|x=0\rangle\;.
\end{equation}
The displacement property thus relates all the states $|x\rangle$
to a particular state $|x=0\rangle$, which remains arbitrary.

We can now ask whether it is possible to satisfy the completeness
property~(\ref{Extwo}) by noting that
\begin{equation}
\int{dx\over C}\,|x\rangle\langle x|=
\sum_{h,h'}\int{dx\over C}\,e^{-ix(h-h')}
\left(\sum_{\alpha,\alpha'}|h,\alpha\rangle\langle h,\alpha|x=0\rangle
\langle x=0|h',\alpha'\rangle\langle h',\alpha'|\right)\;.
\label{intcomp}
\end{equation}
One can arrange that
\begin{equation}
\int{dx\over C}\,e^{-i(h-h')}=\delta_{hh'}\;,
\end{equation}
in which case Eq.~(\ref{intcomp}) simplifies to
\begin{equation}
\int{dx\over C}\,|x\rangle\langle x|=
\sum_h\left(\sum_{\alpha,\alpha'}|h,\alpha\rangle\langle h,\alpha|x=0\rangle
\langle x=0|h,\alpha'\rangle\langle h,\alpha'|\right)\;.
\end{equation}
To make this integral equal to the unit operator requires that
\begin{equation}
\langle h,\alpha|x=0\rangle\langle x=0|h,\alpha'\rangle=\delta_{\alpha\alpha'}
\;\;\;\mbox{for all $h$,}
\end{equation}
which can only be satisfied if the spectrum of $\hat h$ has no degeneracies.
Thus only for non-degenerate $\hat h$ can one hope to find a global optimal
measurement in terms of a POVM described by a single real number $x$.  An
example of how to proceed for a degenerate $\hat h$ can be found in
the discussion of time-energy uncertainty relations in Section~\ref{tE}.

We now assume explicitly that the generator $\hat h$ is non-degenerate,
thus allowing us to drop the degeneracy label $\alpha$ from the
preceding equations.  The form of the completeness relation depends on
further properties of the eigenvalue spectrum of $\hat h$.  We
illustrate the procedure here for the case that the non-degenerate
spectrum of $\hat h$ is discrete (nowhere dense) and that the unitary
generator $e^{-iX\hat h}$ is periodic with smallest period ${\cal X}$,
i.e., $e^{-i{\cal X}\hat h}=\hat1$ (other non-degenerate eigenvalue
spectra are dealt with in the examples of Section~\ref{exgenup}).
This means that all the eigenvalues can be written as
\begin{equation}
h=2\pi n_h/{\cal X}\;,\;\;\mbox{$n_h$ an integer;}
\end{equation}
any discrete spectrum can be approximated in this way for ${\cal X}$
sufficiently large.  The periodicity allows us to restrict both the
parameter $X$ and the measurement results $x$ to the finite interval
$[-{\cal X}/2,{\cal X}/2)$.

The completeness condition~(\ref{Extwo}) now becomes
\begin{eqnarray}
\hat1=
\int_{-{\cal X}/2}^{{\cal X}/2}{dx\over C}\,|x\rangle\langle x|&=&
\sum_{h,h'}|h\rangle\langle h|x=0\rangle\langle x=0|h'\rangle\langle h'|
\int_{-{\cal X}/2}^{{\cal X}/2}{dx\over C}\,e^{-ix(h-h')}\nonumber\\
&=&\sum_{h}|h\rangle\langle h|{{\cal X}|\langle h|x=0\rangle|^2\over C}\;,
\end{eqnarray}
which can be satisfied by choosing $C={\cal X}$ and
\begin{equation}
\langle h|x=0\rangle=e^{if(h)}\;,
\end{equation}
where $f(h)$ is an arbitrary real-valued function.  The completeness
property thus requires that $|x=0\rangle$ have the same magnitude of
overlap with all the eigenstates of $\hat h$.

The minimal choice, $f(h)=0$, which we distinguish by underlining, leads
to {\it canonical\/} states
\begin{equation}
\underline{|x\rangle}=\sum_h|h\rangle e^{-ixh}\;,
\end{equation}
whereas an arbitrary choice for $f(h)$ leads to states,
\begin{equation}
|x\rangle=\sum_h|h\rangle e^{if(h)}e^{-ixh}=
e^{if(\hat h)}\underline{|x\rangle}\;,
\end{equation}
that are a gauge transformation~(\ref{gaugetrans}) of the canonical
states $\underline{|x\rangle}$.  A gauge transformation corresponds
to the freedom to re-phase independently each of the  eigenstates of
$\hat h$---i.e., to replace $|h\rangle$ by $e^{if(h)}|h\rangle$.

The inner product of $|x\rangle$ and $|x'\rangle$ is given by
\begin{equation}
\langle x|x'\rangle=\sum_h e^{i(x-x')h}=
\sum_h e^{2\pi in_h(x-x')/{\cal X}}\;.
\end{equation}
These states are orthogonal---i.e., they can be given $\delta$ function
normalization with $\langle x|x'\rangle={\cal X}\delta(x-x')$---if and
only if {\it all\/} integers are required to represent the eigenvalue
spectrum of $\hat h$; only if the states are orthogonal---i.e.,
all integers are present in the eigenvalue spectrum---are they
eigenstates of a Hermitian operator.

The $x$ and $h$ representations of a state $|\psi\rangle$ are related by
\begin{equation}
{1\over\sqrt{{\cal X}}}\langle x|\psi\rangle=\psi(x)=
{1\over\sqrt{{\cal X}}}\sum_h e^{ixh}e^{-if(h)}\langle h|\psi\rangle\;,
\label{psixh}
\end{equation}
\begin{equation}
e^{-if(h)}\langle h|\psi\rangle={1\over\sqrt{{\cal X}}}
\int_{-{\cal X}/2}^{{\cal X}/2}dx\,e^{-ixh}\psi(x)\;.
\label{psihx}
\end{equation}
The amplitude $e^{-if(h)}\langle h|\psi\rangle$ is the discrete
Fourier coefficient, corresponding to integer $n_h$, of the function
$\psi(x)$, which is periodic with period ${\cal X}$.  The wave functions
$\psi(x)$ are restricted to periodic functions that have vanishing
Fourier coefficients for the unused integers.  By the same token, the
expansion of a  state $|\psi\rangle$ in terms of the states $|x\rangle$,
\begin{equation}
|\psi\rangle=
\int_{-{\cal X}/2}^{{\cal X}/2}
{dx\over{\cal X}}\,|x\rangle\langle x|\psi\rangle=
{1\over\sqrt{{\cal X}}}\int_{-{\cal X}/2}^{{\cal X}/2}dx\,\psi(x)|x\rangle\;,
\end{equation}
is not unique; one can add to $\psi(x)$ any periodic function $g(x)$
that has nonvanishing Fourier coefficients only for the unused integers,
for such a function satisfies
\begin{equation}
{1\over\sqrt{{\cal X}}}\int_{-{\cal X}/2}^{{\cal X}/2}dx\,g(x)|x\rangle=0\;.
\end{equation}
This lack of uniqueness expresses the {\it over\/}completeness of the
states $|x\rangle$.  Both the overcompleteness and the restrictions
on the wave functions $\psi(x)$ are consequences of the lack of
orthogonality of the states $|x\rangle$.

For $H=2\pi n_H/{\cal X}$, we can define a ``displacement operator''
\begin{eqnarray}
\hat D(H)\equiv
\int_{-{\cal X}/2}^{{\cal X}/2}dx\,e^{ixH}\hat E(x)&=&
\int_{-{\cal X}/2}^{{\cal X}/2}{dx\over{\cal X}}\,
e^{ixH}|x\rangle\langle x|\nonumber\\
&=&
\sum_{h,h'}\delta_{n_{h'},n_h+n_H}e^{if(h')}|h'\rangle\langle h|e^{-if(h)}\;,
\label{specdisplacement}
\end{eqnarray}
which displaces eigenstates of $\hat h$; i.e.,
\begin{equation}
\hat D(H)e^{if(h)}|h\rangle=e^{if(h+H)}|h+H\rangle\;,
\end{equation}
provided $H$ is the difference in eigenvalues.  Given a choice of
phases for the eigenstates $|h\rangle$, the canonical states
$\underline{|x\rangle}$ are unique in that their displacement operator
$\underline{\hat D}(H)$ displaces the eigenstates $|h\rangle$ without
the inclusion of any phase factors.  Notice that generally $\hat D(H)$
is not a unitary operator.  For particular eigenvalue spectra of
$\hat h$, however, as in the examples of Section~\ref{exgenup}, the
displacement operator acquires additional important properties.

The $x$ and $h$ representations~(\ref{psixh}) and (\ref{psihx}) of a state
$|\psi\rangle$ show that the condition~(\ref{optwf}) for a global
optimal measurement, when written in the $h$ representation, with
$h=\langle\hat h\rangle+u$, becomes
\begin{equation}
e^{-if(\langle h\rangle+u)}
\Bigl\langle\langle\hat h\rangle+u\!\Bigm|\!\psi_0\Bigr\rangle=
e^{if(\langle h\rangle-u)}
\Bigl\langle\langle\hat h\rangle-u\!\Bigm|\!\psi_0\Bigr\rangle^\ast\;.
\label{opthrep}
\end{equation}
Since the phases in the $h$ representation can be removed by appropriate
choice of the function $f(h)$, this condition reduces to
\begin{equation}
\Bigl|\Bigl\langle\langle\hat h\rangle+u\!\Bigm|\!\psi_0\Bigr\rangle\Bigr|^2=
\Bigl|\Bigl\langle\langle\hat h\rangle-u\!\Bigm|\!\psi_0\Bigr\rangle\Bigr|^2\;.
\label{opthreptwo}
\end{equation}
To make this condition meaningful requires that whenever
$\Bigl\langle\langle\hat h\rangle+u\!\Bigm|\!\psi_0\Bigr\rangle$ is non-zero,
$\langle\hat h\rangle-u$ is an eigenvalue of $\hat h$.  For general
eigenvalue spectra of $\hat h$, the condition~(\ref{opthreptwo}) can be
met by only a very limited class of states, since it requires symmetric
excitation of eigenstates $|h\rangle$ symmetrically located about the
expectation value of $\hat h$.

\section{Examples of Generalized Uncertainty Relations}
\label{exgenup}

We turn now to examples of generalized uncertainty relations, first
dealing, in this section, with nonrelativistic examples and then turning,
in Section~\ref{lorentzup}, to Lorentz-invariant versions of uncertainty
relations.

\subsection{Spatial displacement and momentum}
\label{xp}

The first example of a nonrelativistic uncertainty relation is the
familiar one of spatial displacements $X$ that are generated by
the momentum operator $\hat p$, i.e., $\hat h=\hat p/\hbar$:
\begin{equation}
|\psi_X\rangle=e^{-iX\hat p/\hbar}|\psi_0\rangle\;.
\end{equation}
The uncertainty relation~(\ref{pureup}) takes the form
\begin{equation}
\langle(\delta X)^2\rangle_X\langle(\Delta\hat p)^2\rangle\ge
{\hbar^2\over 4N}\;.
\label{xpup}
\end{equation}
Helstrom \cite{Helstrom} and Holevo [5(Chap.~VI.2)] have presented
parameter-based uncertainty relations for spatial displacement and
momentum, and Dembo, Cover, and Thomas \cite{Dembo} have reviewed
the basis for such uncertainty relations in the properties of
Fisher information.

To investigate the possibilities for optimal POVMs
$\hat E(x)\,dx=|x\rangle\langle x|\,dx/C$, start from the complete set
of $\delta$ function normalized eigenstates $|p\rangle$ of $\hat p$:
\begin{equation}
\langle p|p'\rangle=2\pi\hbar\delta(p-p')\;,
\label{peigenstatesnorm}
\end{equation}
\begin{equation}
\hat1=\int_{-\infty}^\infty{dp\over2\pi\hbar}\,|p\rangle\langle p|\;.
\label{peigenstatescomp}
\end{equation}
The displacement condition~(\ref{Exfour}), with $x=0$ and $X=x$,
becomes
\begin{equation}
\langle p|x\rangle=e^{-ixp/\hbar}\langle p|x=0\rangle\;,
\end{equation}
which leads to
\begin{equation}
\langle p|\!\left(\int_{-\infty}^\infty
{dx\over C}\,|x\rangle\langle x|\right)\!|p'\rangle=
{2\pi\hbar|\langle p|x=0\rangle|^2\over C}\delta(p-p')\;.
\end{equation}
Thus the completeness condition~(\ref{Extwo}) can be satisfied by
choosing $C=1$ and
\begin{equation}
\langle p|x=0\rangle=e^{if(p)}\;,
\end{equation}
where $f(p)$ is an arbitrary real-valued function.  In this case,
because the spectrum of $\hat p$ covers the entire real line, the
states
\begin{equation}
|x\rangle=\int_{-\infty}^\infty{dp\over2\pi\hbar}\,
|p\rangle e^{if(p)}e^{-ixp/\hbar}
\label{genxstates}
\end{equation}
have $\delta$ function normalization,
\begin{equation}
\langle x|x'\rangle=\delta(x-x')\;,
\end{equation}
and thus are eigenstates of the Hermitian operator
\begin{equation}
\hat x=\int_{-\infty}^\infty dx\,x\hat E(x)
=\int_{-\infty}^\infty dx\,x|x\rangle\langle x|\;.
\label{genxop}
\end{equation}

The minimal choice, $f(p)=0$, leads to the canonical position
states,
\begin{equation}
\underline{|x\rangle}=\int_{-\infty}^\infty{dp\over2\pi\hbar}\,
|p\rangle e^{-ixp/\hbar}\;,
\end{equation}
which are eigenstates of the canonical position operator
\begin{equation}
\underline{\hat x}=\int_{-\infty}^\infty dx\,x\underline{\hat E}(x)
=\int_{-\infty}^\infty dx\,x\underline{|x\rangle}\,\underline{\langle x|}\;.
\end{equation}
Measurements described by $\underline{\hat E}(x)$ are thus canonical
measurements of position.  An arbitrary choice for $f(p)$ leads to
the states~(\ref{genxstates}), which, written as
\begin{equation}
|x\rangle=e^{if(\hat p)}\underline{|x\rangle}\;,
\end{equation}
are seen to be a gauge transformation~(\ref{gaugetrans}) of the
position eigenstates.  The state $|x\rangle$ is an eigenstate, with
eigenvalue $x$, of the operator~(\ref{genxop}), which can be written as
\begin{equation}
\hat x=e^{if(\hat p)}\underline{\hat x}e^{-if(\hat p)}=
\underline{\hat x}+\hbar f'(\hat p)\;;
\label{xhat}
\end{equation}
measurements described by $\hat E(x)$ are measurements of this
operator.  Notice that $\hat x$ and $\hat p$ satisfy the canonical
commutation relation,
\begin{equation}
[\hat x,\hat p]=i\hbar\;,
\end{equation}
the gauge freedom being precisely the freedom permitted by this commutator.

The operator that displaces momentum eigenstates,
\begin{equation}
\hat D(P)=\int_{-\infty}^\infty dx\,e^{ixP/\hbar}|x\rangle\langle x|=
e^{i\hat x P/\hbar}
\end{equation}
[cf.~Eq.~(\ref{specdisplacement})], in this case a unitary operator,
acts according to
\begin{equation}
\hat D(P)e^{if(p)}|p\rangle=e^{if(p+P)}|p+P\rangle\;.
\end{equation}
The canonical states $\underline{|x\rangle}$ lead to a displacement
operator $\underline{\hat D}(P)$ that displaces the momentum eigenstates
$|p\rangle$ without the inclusion of any phase factors.

Writing the position wave function of the fiducial state as
$\psi_0(x)=r(x)e^{i\Theta(x)}$, the general relations~(\ref{meanh})
and(\ref{varianceh}) for the mean and variance of $\hat h$
become in this case
\begin{equation}
\langle\hat p\rangle=\int_{-\infty}^\infty dx\,\psi_0^\ast(x)
{\hbar\over i}{\partial\over\partial x}\psi_0(x)=
\int_{-\infty}^\infty dx\,p(x)\hbar\Theta'(x)\;,
\label{meanp}
\end{equation}
\begin{eqnarray}
\langle(\Delta\hat p)^2\rangle&=&
\int_{-\infty}^\infty dx\,\left|
\left(\hbar{\partial\over\partial x}
-i\langle\hat p\rangle\!\right)\!\psi_0(x)\right|^2\nonumber\\
&=&{\hbar^2\over 4}\int_{-\infty}^\infty dx\,{[\,p'(x)]^2\over p(x)}+
\int_{-\infty}^\infty dx\,p(x)[\hbar\Theta'(x)-\langle\hat p\rangle]^2\;.
\label{variancep}
\end{eqnarray}
For the minimal choice [$f(p)=0$] and its canonical position
operator, several authors have drawn attention to the way the
momentum variance splits into the sum of the two parts in
Eq.~(\ref{variancep}).  Stam \cite{Stam} noted long ago that
the variance of $\hat p$ is bounded below by the Fisher information
for position measurements, Cohen \cite{Cohen} has discussed
and illustrated with examples the split of the momentum
variance, and Sipe and Arkani-Hamed \cite{Sipe} have used this
split and the similar split of the variance of $\underline{\hat x}$
to contrast the coherence of pure and mixed states.

The condition for a global optimal measurement is that the position
wave function of the fiducial state have the form
\begin{equation}
\langle x|\psi_0\rangle=\psi_0(x)=r(x)e^{i\langle\hat p\rangle x/\hbar}
\end{equation}
[cf.~Eq.~(\ref{optwf})].  Transforming to the momentum representation,
with $p=\langle p\rangle+u$,
\begin{equation}
e^{-if(\langle\hat p\rangle+u)}
\Bigl\langle\langle\hat p\rangle+u\!\Bigm|\!\psi_0\Bigr\rangle=
\int_{-\infty}^\infty dx\,e^{-ixu/\hbar}r(x)\;,
\end{equation}
one sees that the optimality condition can be written as
\begin{equation}
e^{-if(\langle\hat p\rangle+u)}
\Bigl\langle\langle\hat p\rangle+u\!\Bigm|\!\psi_0\Bigr\rangle=
e^{if(\langle\hat p\rangle-u)}
\Bigl\langle\langle\hat p\rangle-u\!\Bigm|\!\psi_0\Bigr\rangle^\ast
\label{xpopt}
\end{equation}
[cf.~Eq.~(\ref{opthrep})].  If one is restricted to canonical position
measurements, for which $f(p)=0$, the condition for optimality is that
\begin{equation}
\Bigl\langle\langle\hat p\rangle+u\!\Bigm|\!\psi_0\Bigr\rangle=
\Bigl\langle\langle\hat p\rangle-u\!\Bigm|\!\psi_0\Bigr\rangle^\ast\;.
\label{xpoptone}
\end{equation}
If one allows gauge-transformed measurements, then the gauge
transformation can be used to remove the phases in the momentum
representation, so the condition for optimality is the weaker
condition that
\begin{equation}
\Bigl|\Bigl\langle\langle\hat p\rangle+u\!\Bigm|\!\psi_0\Bigr\rangle\Bigr|^2=
\Bigl|\Bigl\langle\langle\hat p\rangle-u\!\Bigm|\!\psi_0\Bigr\rangle\Bigr|^2\;,
\label{xpopttwo}
\end{equation}
i.e., that the momentum probability density is symmetric about
$\langle\hat p\rangle$.

It is instructive to illustrate these ideas with an extended example
based on a specific fiducial state.  For this purpose, introduce an
``annihilation operator''
\begin{equation}
\hat a={1\over\sqrt2}
\left({\underline{\hat x}\over L}+i{L\hat p\over\hbar}\right)\;,
\end{equation}
where $L$ is a constant that has dimensions of length, and a ``vacuum
state'' $|{\rm vac}\rangle$, which is the state annihilated by $\hat a$,
\begin{equation}
\hat a|{\rm vac}\rangle=0\;.
\label{annvac}
\end{equation}
One easily verifies from this equation that in the vacuum state,
$\underline{\hat x}$ and $\hat p$ have zero mean, and their covariance
matrix is given by
\begin{equation}
{\langle{\rm vac}|(\Delta\underline{\hat x})^2|{\rm vac}\rangle\over L^2}=
{L^2\langle{\rm vac}|(\Delta\hat p)^2|{\rm
vac}\rangle\over\hbar^2}={1\over2}\;,
\end{equation}
\begin{equation}
\langle{\rm vac}|(\Delta\underline{\hat x}\Delta\hat p
+\Delta\hat p\Delta\underline{\hat x})|{\rm vac}\rangle=0\;.
\end{equation}
The vacuum state is thus a minimum-uncertainty state for
$\underline{\hat x}$ and $\hat p$.  It is convenient throughout
the remainder of this example to use units such that $L=1$, a
choice that gives $\underline{\hat x}$ and $\hat p/\hbar$ equal
variances in the vacuum state.

The next step is to introduce the squeeze operator \cite{Schumaker}
\begin{equation}
\hat S\equiv
\exp\!\left({1\over 2}r\left(e^{-2i\varphi}\hat a^2-
e^{2i\varphi}\hat a^\dagger\mbox{}^2\right)\right)\;,
\end{equation}
which is a function of a squeeze parameter $r\ge0$ and a squeeze angle
$\varphi$.  The squeeze operator has the property \cite{Schumaker}
\begin{eqnarray}
\hat S\hat a\hat S^\dagger&=&
\hat a\cosh r+\hat a^\dagger e^{2i\varphi}\sinh r\nonumber\\
&=&{1\over\sqrt2}\left(\underline{\hat x}(\cosh r+e^{2i\varphi}\sinh r)+
i{\hat p\over\hbar}(\cosh r-e^{2i\varphi}\sinh r)\right)
\equiv\hat\chi\;.
\label{Sprop}
\end{eqnarray}
The fiducial state in this example is generated from the vacuum state
by the squeeze operator,
\begin{equation}
|\psi_0\rangle=\hat S|{\rm vac}\rangle\;,
\label{squeezedvac}
\end{equation}
and is sometimes called the squeezed vacuum state.  An immediate
consequence of the property~(\ref{Sprop}) is that the squeezed
vacuum state is annihilated by $\hat\chi$:
\begin{equation}
\hat\chi|\psi_0\rangle=0\;.
\label{annsqueezedvac}
\end{equation}

One can get a better feel for the nature of the squeezed vacuum
state and, in particular, its parameters $r$ and $\varphi$ by
considering $\underline{\hat x}$ and $\hat p/\hbar$ to be co\"ordinates on
a phase plane and then rotating by angle $\varphi$ to new canonical
co\"ordinates $\underline{\hat x}'$ and $\hat p'/\hbar$:
\begin{equation}
{1\over\sqrt2}(\underline{\hat x}+i\hat p/\hbar)=
\hat a=\hat a'e^{i\varphi}=
{1\over\sqrt2}(\underline{\hat x}'+i\hat p'/\hbar)e^{i\varphi}\;.
\end{equation}
In terms of the rotated co\"ordinates the operator $\hat\chi$ assumes
the form
\begin{equation}
\hat\chi=e^{i\varphi}(\hat a'\cosh r+\hat a'\mbox{}^\dagger\sinh r)=
{1\over\sqrt2}e^{i\varphi}
\left(\underline{\hat x}'e^r+i{\hat p'\over\hbar}e^{-r}\right)\;,
\end{equation}
which, together with Eq.~(\ref{annsqueezedvac}), implies that in the
squeezed vacuum state, $\underline{\hat x}'$ and $\hat p'$ have
zero mean, and their covariance matrix is given by
\begin{equation}
\langle\psi_0|(\Delta\underline{\hat x}')^2|\psi_0\rangle e^{2r}=
{\langle\psi_0|(\Delta\hat p')^2|\psi_0\rangle\over\hbar^2}e^{-2r}=
{1\over2}\;,
\end{equation}
\begin{equation}
\langle\psi_0|(\Delta\underline{\hat x}'\Delta\hat p'
+\Delta\hat p'\Delta\underline{\hat x}')|\psi_0\rangle=0\;.
\end{equation}
The squeezed vacuum state is thus a minimum-uncertainty state for
the rotated co\"ordinates $\underline{\hat x}'$ and $\hat p'$; relative
to the vacuum state, $\underline{\hat x}'$ has uncertainty reduced
by a factor $e^{-r}$, and $\hat p'$ has uncertainty increased by a
factor $e^r$.  Figure~\ref{onlyfig} depicts the squeezed vacuum
state on a phase-plane diagram.

If one rotates to any other orthogonal axes, the position variance
gets bigger than the variance of $\underline{\hat x}'$ (recall that
$r\ge0$), because the reduced variance of $\underline{\hat x}'$ is
contaminated by the increased variance of $\hat p'$.  Indeed, the
covariance matrix of the original canonical co\"ordinates, obtained
directly from Eq.~(\ref{annsqueezedvac}) or by rotating back to the
original co\"ordinates, is given by \cite{Schumaker}
\begin{equation}
\langle\psi_0|(\Delta\underline{\hat x})^2|\psi_0\rangle=
{1\over2}\left(e^{-2r}\cos^2\!\varphi+e^{2r}\sin^2\!\varphi\right)=
{1\over2{\rm Re}(\gamma)}\;,
\end{equation}
\begin{equation}
{\langle\psi_0|(\Delta\hat p)^2|\psi_0\rangle\over\hbar^2}=
{1\over2}\left(e^{-2r}\sin^2\!\varphi+e^{2r}\cos^2\!\varphi\right)=
{1\over2{\rm Re}(\gamma^{-1})}\;,
\end{equation}
\begin{equation}
{{1\over2}\langle\psi_0|(\Delta\underline{\hat x}\Delta\hat p
+\Delta\hat p\Delta\underline{\hat x})|\psi_0\rangle\over\hbar}=
-{1\over2}\sinh 2r\sin 2\varphi=
-{{\rm Im}(\gamma)\over2{\rm Re}(\gamma)}=
{{\rm Im}(\gamma^{-1})\over2{\rm Re}(\gamma^{-1})}\;,
\end{equation}
where
\begin{equation}
\gamma={\cosh r+e^{2i\varphi}\sinh r\over\cosh r-e^{2i\varphi}\sinh r}=
{1+i\sinh2r\sin2\varphi\over\cosh2r-\sinh2r\cos2\varphi}=
{\cosh2r+\sinh2r\cos2\varphi\over1-i\sinh2r\sin2\varphi}
\end{equation}
is a complex constant.  This covariance matrix can also be gotten
from the wave function of the fiducial state in the canonical position
representation \cite{Schumaker},
\begin{equation}
\underline{\psi_0}(x)=\underline{\langle x|}\psi_0\rangle=
\left({{\rm Re}(\gamma)\over\pi}\right)^{1/4}
\exp\!\left(-{\gamma x^2\over2}\right)\;,
\label{standxwf}
\end{equation}
which follows from integrating the differential equation that
represents Eq.~(\ref{annsqueezedvac}) in the canonical position
representation.  An irrelevant phase factor is set equal to
unity in the wave function~(\ref{standxwf}) (Schumaker \cite{Schumaker}
has given a consistent set of phases for squeezed-state wave functions).

It is now straightforward to find the optimal measurement.  The wave
function in the momentum basis is given by
\begin{equation}
\langle p|\psi_0\rangle=
\int_{-\infty}^\infty dx\,e^{-ixp/\hbar}\underline{\psi_0}(x)=
\sqrt{{|\gamma|\over\gamma}}
\left(4\pi{\rm Re}(\gamma^{-1})\right)^{1/4}
\exp\!\left(-{p^2/\hbar^2\over2\gamma}\right)\;,
\end{equation}
where $\sqrt{|\gamma|/\gamma}$ is an overall phase factor.
According to the optimality condition~(\ref{xpopt}), choosing $f(p)$
to cancel the imaginary part of this complex Gaussian, i.e.,
\begin{equation}
f(p)=-{1\over 2}{\rm Im}(\gamma^{-1})p^2/\hbar^2\;,
\end{equation}
yields an optimal measurement, corresponding to measuring the operator
\begin{equation}
\hat x=\underline{\hat x}+\hbar f'(\hat p)=
\underline{\hat x}-{\rm Im}(\gamma^{-1})\hat p/\hbar\;.
\label{optop}
\end{equation}
The distinguishing feature of using a squeezed state as the fiducial
state is that the optimal measurement is a {\it linear\/} combination
of $\underline{\hat x}$ and $\hat p$.  Transforming to the $x$
representation yields a real wave function
\begin{equation}
\psi_0(x)=\langle x|\psi_0\rangle=\sqrt{{|\gamma|\over\gamma}}
\left({1\over\pi{\rm Re}(\gamma^{-1})}\right)^{1/4}
\exp\!\left(-{x^2\over2{\rm Re}(\gamma^{-1})}\right)\;,
\end{equation}
aside from the overall phase factor $\sqrt{|\gamma|/\gamma}$, in
accordance with the general condition for an optimal measurement.

One feature of the optimal measurement in this case, which follows
from the fact that $\psi_0(x)$ is a Gaussian wave function of the sort
considered at the end of Section~\ref{optmeasA}, deserves emphasis.
The probability density of measurements of $\hat x$,
\begin{equation}
p(x)=|\psi_0(x)|^2=
\left({1\over\pi{\rm Re}(\gamma^{-1})}\right)^{1/2}
\exp\!\left(-{x^2\over{\rm Re}(\gamma^{-1})}\right)\;,
\label{optprobx}
\end{equation}
is a zero-mean Gaussian with variance
\begin{equation}
\langle\psi_0|(\Delta\hat x)^2|\psi_0\rangle=
{1\over2}{\rm Re}(\gamma^{-1})=
{1\over2}\left(e^{-2r}\sin^2\!\varphi+e^{2r}\cos^2\!\varphi\right)^{-1}
={\hbar^2\over4\langle\psi_0|(\Delta\hat p)^2|\psi_0\rangle}\;.
\label{optvarx}
\end{equation}
Generally one must appeal to the large-$N$ asymptotic limit to saturate
the first (classical) inequality in Eq.~(\ref{BCbound})---i.e., to
achieve the Cram\'er-Rao bound---but since the statistics of $\hat x$
are Gaussian, no such appeal is necessary.  Indeed, for Gaussian
statistics the sample mean~(\ref{samplemean}) of the data (here
$\langle x\rangle_0=0$) provides an efficient estimator for all values
of $N$, as is discussed at the end of Section~\ref{optmeasA}.  The
Gaussian statistics of $x$ for the fiducial state, displaced according to
Eq.~(\ref{pxX}), imply that $\langle X_{\rm est}\rangle_X=X$---i.e.,
the estimator is unbiased---which means that the estimate's deviation
away from the parameter becomes $\delta X=X_{\rm est}-X=\Delta X_{\rm est}$.
The mean-square deviation is independent of $X$ and reduces to
\begin{equation}
\langle(\delta X)^2\rangle=\langle(\Delta X_{\rm est})^2\rangle=
{1\over N}\langle\psi_0|(\Delta\hat x)^2|\psi_0\rangle={1\over NF}\;.
\label{deltaXex}
\end{equation}
The final equality, which shows that $X_{\rm est}$ is an efficient
estimator, follows most easily from the form of the Fisher information
in Eq.~(\ref{Fisherinfotwo}).  In this case, where an efficient estimator
is known, one can proceed directly to equality in the uncertainty
relation~(\ref{xpup}), without going through the Fisher information,
by combining Eqs.~(\ref{optvarx}) and (\ref{deltaXex}).

One gains insight into the optimal measurement by writing the
measured operator~(\ref{optop}) as
\begin{equation}
\hat x=\underline{\hat x}+{\hat p\over\hbar}\tan\theta=
{\underline{\hat x}\cos\theta+(\hat p/\hbar)\sin\theta\over\cos\theta}\;,
\label{optop2}
\end{equation}
where
\begin{equation}
\tan\theta=-{\rm Im}(\gamma^{-1})
={\sinh2r\sin2\varphi\over\cosh 2r+\sinh2r\cos2\varphi}\;,
\end{equation}
and regarding $\hat x$ as a species of position operator that arises
from a rotation in the phase plane by angle $\theta$, followed by
rescaling by $1/\cos\theta$.  The rescaling means that displacement
by $X$ produces the same ``signal'' in $\hat x$ as it does in
$\underline{\hat x}$.  The optimal angle $\theta$ is not equal
to $\varphi$, the rotation angle that minimizes the variance of
the rotated position; instead, the optimal angle is a compromise
between reduced ``noise'' and reduced signal, both of which come with
rotation (see Fig.~\ref{onlyfig}).  The rescaling of $\hat x$ accounts
for the reduced signal, so the variance of $\hat x$,
\begin{eqnarray}
&\mbox{}&\!\!\langle\psi_0|(\Delta\hat x)^2|\psi_0\rangle\nonumber\\
&\mbox{}&\;\mbox{}=
\langle\psi_0|(\Delta\underline{\hat x})^2|\psi_0\rangle+
2{{1\over2}\langle\psi_0|(\Delta\underline{\hat x}\Delta\hat p
+\Delta\hat p\Delta\underline{\hat x})|\psi_0\rangle\over\hbar}\tan\theta+
{\langle\psi_0|(\Delta\hat p)^2|\psi_0\rangle\over\hbar^2}\tan^2\theta\;,\;\;\;
\nonumber\\
&&
\label{nsr}
\end{eqnarray}
is a noise-to-signal ratio \cite{Braunstein}.  Indeed, the angle that
minimizes this noise-to-signal ratio,
\begin{equation}
\tan\theta=
-{{1\over2}\langle\psi_0|(\Delta\underline{\hat x}\Delta\hat p
+\Delta\hat p\Delta\underline{\hat x})|\psi_0\rangle/\hbar\over
\langle\psi_0|(\Delta\hat p)^2|\psi_0\rangle/\hbar^2}=
-{\rm Im}(\gamma^{-1})\;,
\label{minnsr}
\end{equation}
defines the optimal measurement.

\subsection{Harmonic-oscillator phase and number of quanta}
\label{phin}

For our second example of a nonrelativistic uncertainty relation,
consider a harmonic oscillator that has creation and annihilation
operators $\hat a^\dagger$ and $\hat a$.  The ``number operator''
\begin{equation}
\hat n=\hat a^\dagger\hat a
\end{equation}
has eigenstates $|n\rangle$, called ``number states,'' where $n=0,1,\ldots$
is the number of quanta.  Shifts $X=\Phi$ in the phase of the
oscillator are generated by the number operator,
\begin{equation}
|\psi_\Phi\rangle=e^{i\Phi\hat n}|\psi_0\rangle\;,
\end{equation}
i.e., $\hat h=-\hat n$.  The uncertainty relation~(\ref{pureup})
then reads
\begin{equation}
\langle(\delta \Phi)^2\rangle_\Phi\langle(\Delta\hat n)^2\rangle
\ge{1\over4N}\;.
\label{phinup}
\end{equation}
Holevo [5(Chap.~III.9)] has considered this sort of phase uncertainty
relation.  Lane, Braunstein, and Caves \cite{Lane}, in a detailed
analysis of phase measurements, have used the formula~(\ref{varianceh}),
specialized to give the variance of the number operator, to bound
the Fisher information for the phase.

A phase shift $\Phi$ can be thought of as a dimensionless time
[measured in units of (harmonic-oscillator period)/$2\pi$], so
the uncertainty relation~(\ref{phinup}) is a dimensionless
time-energy uncertainty relation, special because of the uniform
spacing of the eigenstates of the generator $\hat n$.  General
time-energy uncertainty relations, corresponding to other energy
spectra, are considered in Section~\ref{tE}.

Since phase shifts are periodic with period $2\pi$, $\Phi$ can be
restricted to the interval $-\pi\le\Phi<\pi$.  It might be thought
that there is a difficulty with the phase-number uncertainty
relation~(\ref{phinup}) when the fiducial state is a number state,
for which $\langle(\Delta\hat n)^2\rangle=0$; the uncertainty relation
then forces $\langle(\delta\Phi)^2\rangle_\Phi\rightarrow\infty$, even
though a sensible estimator $\Phi_{\rm est}$ is restricted to
the same $2\pi$ interval as is $\Phi$.  No difficulty arises,
however, because for a number state, no measurement can provide
any information about the phase shift; thus, any estimator, sensible
or not, satisfies $d\langle\Phi_{\rm est}\rangle/d\Phi=0$, with
the result that the deviation $\delta\Phi$ of Eq.~(\ref{deltaX})
diverges, even if $\Phi_{\rm est}$ is restricted to a finite range.

The possibilities for POVMs
$\hat E(\phi)\,d\phi=|\phi\rangle\langle\phi|\,d\phi/C$ ($-\pi\le\phi<\pi$)
are determined by the displacement condition~(\ref{Exfour}),
which, with $\phi=0$ and $\Phi=\phi$, becomes
\begin{equation}
\langle n|\phi\rangle=e^{i\phi n}\langle n|\phi=0\rangle\;,
\end{equation}
and by the completeness condition~(\ref{Extwo}),
\begin{equation}
\hat1=\int_{-\pi}^\pi{d\phi\over C}\,|\phi\rangle\langle\phi|=
{2\pi\over C}\sum_{n=0}^\infty
|\langle n|\phi=0\rangle|^2\,|n\rangle\langle n|\;,
\end{equation}
which can be satisfied by choosing $C=2\pi$ and
\begin{equation}
\langle n|\phi=0\rangle=e^{if(n)}\;,
\end{equation}
where $f(n)$ is an arbitrary real-valued function.  Since there are
no number states for negative integers, the phase states $|\phi\rangle$
are not orthogonal, the inner product being given by \cite{ShapiroShepard}
\begin{eqnarray}
\langle\phi|\phi'\rangle&=&
\sum_{n=0}^\infty e^{-i(\phi-\phi')n}\nonumber\\
&=&{1\over2}\!\left(\sum_{n=-\infty}^\infty e^{-i(\phi-\phi')n}+
\sum_{n=-\infty}^\infty{\rm sgn}(n)e^{-i(\phi-\phi')n}+1\right)\nonumber\\
&=&\pi\delta(\phi-\phi')-{i\over2}\cot\!\left({\phi-\phi'\over2}\right)
+{1\over2}\;,
\end{eqnarray}
where
\begin{equation}
{\rm sgn}(n)\equiv
\left\{\begin{array}{ll}
         -1\;,\;\;&\mbox{$n<0$,}\\
         0\;,\;\;&\mbox{$n=0$,}\\
         1\;,\;\;&\mbox{$n>0$.}\\
       \end{array}
\right.
\label{sgn}
\end{equation}
Hence the states $|\phi\rangle$ are overcomplete and are not the
eigenstates of any Hermitian operator.  There is no Hermitian phase
operator in the infinite-dimensional Hilbert space of a harmonic oscillator
\cite{Holevo,ShapiroShepard,SusskindGlogower,CarruthersNieto,Hall91},
although one can be constructed if the harmonic-oscillator Hilbert
space is truncated to be finite-dimensional \cite{PeggBarnett}.

The minimal choice, $f(n)=0$, leads to the Susskind-Glogower
\cite{SusskindGlogower} canonical phase states,
\begin{equation}
\underline{|\phi\rangle}=\sum_{n=0}^\infty|n\rangle e^{i\phi n}\;,
\end{equation}
which are eigenstates of the non-unitary number-lowering operator
\begin{equation}
\widehat{e^{i\phi}}\equiv
(\hat n+1)^{-1/2}\hat a=\hat a\hat n^{-1/2}
=\sum_{n=1}^\infty|n-1\rangle\langle n|\;,
\end{equation}
i.e.,
\begin{equation}
\widehat{e^{i\phi}}\underline{|\phi\rangle}=
e^{i\phi}\underline{|\phi\rangle}\;.
\end{equation}
Helstrom \cite{Helstrom} and Holevo [5(Chap~III.9)] have considered
measurements described by the Susskind-Glogower states.  An arbitrary
choice for $f(n)$ leads to states,
\begin{equation}
|\phi\rangle=\sum_{n=0}^\infty|n\rangle e^{if(n)}e^{i\phi n}=
e^{if(\hat n)}\underline{|\phi\rangle}\;,
\end{equation}
that are a gauge transformation~(\ref{gaugetrans}) of the
Susskind-Glogower states.  The state $|\phi\rangle$ is an eigenstate,
with eigenvalue $e^{i\phi}$, of the operator
\begin{equation}
e^{if(\hat n)}\widehat{e^{i\phi}}e^{-if(\hat n)}=
e^{-i[f(\hat n+1)-f(\hat n)]}\widehat{e^{i\phi}}=
\widehat{e^{i\phi}}e^{-i[f(\hat n)-f(\hat n-1)]}
=\sum_{n=1}^\infty e^{if(n-1)}|n-1\rangle\langle n|e^{-if(n)}\;;
\end{equation}
the differences $f(\hat n+1)-f(\hat n)$ and $f(\hat n)-f(\hat n-1)$
in the exponents are analogous to the derivative $\hbar f'(\hat p)$
in Eq.~(\ref{xhat}).

For $N$ an integer the number displacement operator is given by
\begin{eqnarray}
\hat D(N)\equiv
\int_{-\pi}^{\pi}d\phi\,e^{-i\phi N}\hat E(\phi)
&=&\int_{-\pi}^{\pi}{d\phi\over2\pi}\,
e^{-i\phi N}|\phi\rangle\langle\phi|\nonumber\\
&=&\sum_{n,n'}\delta_{n',n+N}\,
e^{if(n')}|n'\rangle\langle n|e^{-if(n)}\nonumber\\
&=&\sum_{n=\max(0,-N)}^\infty e^{if(n+N)}|n+N\rangle\langle n|e^{-if(n)}
\end{eqnarray}
[cf.~Eq.~(\ref{specdisplacement})].  Because there are no number states
for negative integers, $\hat D(N)$ is not unitary; the final form of
$\hat D(N)$ is a consequence of the regular spacing of the number states.
Notice that $\hat D(-1)=e^{if(\hat n)}\widehat{e^{i\phi}}e^{-if(\hat n)}$
[thus the states $|\phi\rangle$ are eigenstates of $\hat D(-1)$] and
$\hat D(1)=[\hat D(-1)]^\dagger=
e^{if(\hat n)}\widehat{e^{i\phi}}^\dagger e^{-if(\hat n)}$.

The $\phi$ and $n$ representations of a state $|\psi\rangle$ are related
by
\begin{equation}
{1\over\sqrt{2\pi}}\langle\phi|\psi\rangle=\psi(\phi)=
{1\over\sqrt{2\pi}}\sum_n e^{-i\phi n}e^{-if(n)}\langle n|\psi\rangle\;,
\end{equation}
\begin{equation}
e^{-if(n)}\langle n|\psi\rangle=
{1\over\sqrt{2\pi}}\int_{-\pi}^\pi d\phi\,e^{i\phi n}\psi(\phi)\;,
\end{equation}
$e^{-if(n)}\langle n|\psi\rangle$ being the Fourier coefficient of
the periodic function $\psi(\phi)$.  The condition for a global optimal
measurement, that the $\phi$ wave function of the fiducial state have
the form
\begin{equation}
\psi_0(\phi)=r(\phi)e^{-i\langle\hat n\rangle\phi}\;,
\end{equation}
is equivalent to the following requirement on the number-state amplitudes:
\begin{equation}
e^{-if(\langle n\rangle+u)}
\Bigl\langle\langle\hat n\rangle+u\!\Bigm|\!\psi_0\Bigr\rangle=
e^{if(\langle n\rangle-u)}
\Bigl\langle\langle\hat n\rangle-u\!\Bigm|\!\psi_0\Bigr\rangle^\ast
\end{equation}
[cf.~Eq.~(\ref{opthrep})].  If one is restricted to Susskind-Glogower
phase measurements [$f(n)=0$], the condition for optimality is that
\begin{equation}
\Bigl\langle\langle\hat n\rangle+u\!\Bigm|\!\psi_0\Bigr\rangle=
\Bigl\langle\langle\hat n\rangle-u\!\Bigm|\!\psi_0\Bigr\rangle^\ast\;,
\end{equation}
but if one allows gauge-transformed measurements, the condition for
optimality becomes
\begin{equation}
\Bigl|\Bigl\langle\langle n\rangle+u\!\Bigm|\!\psi_0\Bigr\rangle\Bigr|^2=
\Bigl|\Bigl\langle\langle n\rangle-u\!\Bigm|\!\psi_0\Bigr\rangle\Bigr|^2\;.
\end{equation}
In either case, the condition for optimality can only be met by a
limited class of states; in particular, because of the discreteness of
the number states, $\langle\hat n\rangle$ must be integral or
half-integral, and because of the lower bound at $n=0$,
$\langle n|\psi\rangle$ must vanish for $n>2\langle\hat n\rangle$.

Since the optimality conditions appear to be so restrictive, it is
worth noting that a large class of ``semiclassical'' states satisfy
them approximately.  By a semiclassical state, we mean one that has
number amplitudes $\langle n|\psi\rangle$ that are concentrated at
large $n$, rendering the lower bound at $n=0$ irrelevant, and are
spread over a wide range of values of $n$, making the discreteness of
$n$ unimportant.  For semiclassical states measurements described by
$\hat E(\phi)$ are nearly optimal provided only that the number
probabilities $|\langle n|\psi\rangle|^2$ are symmetric about
$\langle\hat n\rangle$ [cf.~Eq.~(\ref{xpopttwo})].  The extent to
which measurements of $\hat E(\phi)$ are sub-optimal for semiclassical
states deserves further investigation.

\subsection{Time and energy}
\label{tE}

For our final example of a nonrelativistic uncertainty relation, consider
the Hilbert-space path traced out by dynamical evolution under the
Hamiltonian $\hat H$:
\begin{equation}
|\psi_T\rangle=e^{-iT\hat H/\hbar}|\psi_0\rangle \;.
\end{equation}
The parameter here is the elapsed time $T$, and the temporal
displacements are generated by $\hat h=\hat H/\hbar$.  The uncertainty
relation~(\ref{pureup}) reads
\begin{equation}
\langle(\delta T)^2\rangle_T\langle(\Delta\hat H)^2\rangle\ge
{\hbar^2\over4N}\;.
\label{tEup}
\end{equation}
This inequality means that no matter what measurements are made to
determine the elapsed time $T$ and no matter how the data from those
measurements are processed to give an estimate of $T$, the estimator's
mean-square deviation from the actual elapsed time must satisfy
Eq.~(\ref{tEup}).

The time-energy uncertainty relation~(\ref{tEup}) must be used carefully,
however.  For example, suppose one wishes to estimate elapsed time from the
dynamics of a small system decaying into an environment.  The
inequality~(\ref{tEup}) places useful limits on such an estimate only
if one uses the {\it total\/} Hamiltonian of the system and the
environment.  An alternative approach, which focuses on the dissipative
dynamics of the small system, is to use a master equation to describe
the dynamics of the small system, to compute $ds/dT$ from the master
equation, and then to use the original inequality~(\ref{BCbound}) to
place limits on the estimation of elapsed time \cite{BraunsteinMilburn}.

Mandelstam and Tamm \cite{MandelstamTamm} derived the first
parameter-based uncertainty relation, for time and energy, in the
following way.  They realized that to measure elapsed time $T$,
one measures an observable $\hat A$ that changes with time---a clock
observable.  By defining a time uncertainty
\begin{equation}
\Delta T\equiv
{\langle(\Delta\hat A)^2\rangle^{1/2}\over|d\langle\hat A\rangle/dT|}=
{\hbar\langle(\Delta\hat A)^2\rangle^{1/2}\over
|\langle[\hat A,\hat H]\rangle|}\;,
\end{equation}
they converted the standard operator uncertainty relation for $\hat A$
and $\hat H$,
\begin{equation}
\langle(\Delta\hat A)^2\rangle^{1/2}
\langle(\Delta\hat H)^2\rangle^{1/2}\ge
{1\over2}|\langle[\hat A,\hat H]\rangle|\;,
\end{equation}
into a time-energy uncertainty relation
\begin{equation}
\Delta T\langle(\Delta\hat H)^2\rangle^{1/2}\ge{\hbar\over2}\;.
\end{equation}

The key idea in Mandelstam and Tamm's work, to regard elapsed time
as a parameter to be determined by measuring some other quantity,
underlies the formalism of parameter-based uncertainty relations.
The technical advances in the present formalism are, first, the use
of estimation theory to incorporate easily the possibility of multiple
measurements and to quantify precisely the precision with which a
parameter can be determined and, second, the use of POVMs to allow for
all quantum measurements that might be used to infer the parameter.
Helstrom \cite{Helstrom} and Holevo [5(Chaps.~III.8 and IV.7] pioneered
in using these technical advances to formulate time-energy uncertainty
relations.  Hilgevoord and Uffink \cite{Hilgevoord88,Hilgevoord90} have
formulated a different sort of parameter-based time-energy uncertainty
relation.

For the case of pure-state time evolution, Anandan and Aharonov
\cite{AnandanAharonov} noted the connection between the Hilbert-space
angle~(\ref{Hangle}) and the variance of the Hamiltonian $\hat H$.
This connection follows from combining Eqs.~(\ref{s2variance}) and
(\ref{Hangles2}):
\begin{equation}
{d\theta\over dT}={1\over2}{ds\over dT}=
{\langle(\Delta\hat H)^2\rangle_T^{1/2}\over\hbar}\;.
\end{equation}
Knowing that Hilbert-space angle is related to distinguishability
through the inner product, Anandan and Aharonov formulated an uncertainty
relation by asking for the minimum time for the evolution to proceed to
an orthogonal state.  Anandan \cite{Anandan} and Uhlmann \cite{Uhlmann}
generalized this approach to mixed states.  Our formulation differs in
that we also relate Hilbert-space angle to statistical distance and thus
to a precise measure of the uncertainty in determining the elapsed time
$T$, i.e., the minimum mean-square deviation $\langle(\delta T)^2\rangle_X$.

The states $|t\rangle$ that are used to describe global optimal
measurements can be obtained from the energy eigenstates $|\epsilon\rangle$:
\begin{equation}
\hat H|\epsilon\rangle = \epsilon|\epsilon\rangle \;.
\end{equation}
If the spectrum of energy eigenvalues is discrete and non-degenerate,
then the time representation follows immediately from obvious changes
in the notation of Section~\ref{optmeas}.  For example, the time states
are given by
\begin{equation}
|t\rangle=
\sum_\epsilon |\epsilon\rangle e^{if(\epsilon)}e^{-it\epsilon/\hbar}\;,
\label{tstates}
\end{equation}
with the minimal choice, $f(\epsilon)=0$, giving the canonical time
representation.  The states $|t\rangle$, like position eigenstates
and phase states, are generally not physical states, as they typically
have infinite energy.  Holevo [5(Chaps.~III.8 and IV.7)] has considered
the canonical time representation and its application to optimal
measurements and has worked out in detail the example of a free
particle, where the energy spectrum is continuous and doubly degenerate.
We review the free-particle example here to provide an example of
how to proceed when the generator $\hat h$ is degenerate.

Consider then a free particle with Hamiltonian
\begin{equation}
\hat H=\hat p^2/2m\;.
\end{equation}
The energy eigenstates coincide with the momentum eigenstates $|p\rangle$,
which we normalize as in Eq.~(\ref{peigenstatesnorm}).  The energy
eigenstates are, however, doubly degenerate (except for $p=0$), with
eigenvalues given by
\begin{equation}
\epsilon=p^2/2m\;.
\end{equation}
A convenient way to deal with the degeneracy is to introduce a degeneracy
label
\begin{equation}
\sigma={\rm sgn}(p)
\end{equation}
[$|p|=\sigma p$; cf.~Eq.~(\ref{sgn})], which allows us to write
\begin{equation}
p=\sigma\sqrt{2m\epsilon}\;.
\end{equation}
The energy eigenstates can now be defined as
\begin{equation}
|\epsilon,\sigma\rangle=(m/2\epsilon)^{1/4}|p\rangle
\;\;\;\Longleftrightarrow\;\;\;
|p\rangle=(\sigma p/m)^{1/2}|\epsilon,\sigma\rangle\;,
\end{equation}
where $\sigma$ is used to distinguish degenerate energy eigenstates
and where the normalization is chosen so that
\begin{equation}
\langle\epsilon,\sigma|\epsilon',\sigma'\rangle
=2\pi\hbar\delta_{\sigma\sigma'}\delta(\epsilon-\epsilon')\;,
\end{equation}
\begin{equation}
\hat1=\sum_\sigma\int_0^\infty{d\epsilon\over2\pi\hbar}\,
|\epsilon,\sigma\rangle\langle\epsilon,\sigma|\;.
\end{equation}

We can now find global optimal measurements in terms of time states
$|t,\sigma\rangle$, where the states $|t,+1\rangle$ are constructed
as in Eq.~(\ref{tstates}), but in the $\sigma=+1$ subspace of Hilbert
space, and the states $|t,-1\rangle$ are similarly constructed in the
$\sigma=-1$ subspace.  Notice, however, that because of the degeneracy
we have the freedom not only to re-phase each of the energy eigenstates
independently, but also to use as the basic energy eigenstates any
orthonormal linear combination of the states $|\epsilon,+1\rangle$ and
$|\epsilon,-1\rangle$.  In symbols, we have the freedom to choose new
energy eigenstates
\begin{equation}
|\epsilon,\gamma\rangle=
\sum_\sigma|\epsilon,\sigma\rangle e^{if(\epsilon)}
U_{\sigma\gamma}(\epsilon)\;,
\label{egamma}
\end{equation}
where $\gamma=\pm1$ is a new degeneracy label and
$U_{\sigma\gamma}(\epsilon)$ is a $2\times2$ unitary matrix with
unit determinant.

With this freedom in mind, we seek a global optimal measurement in
terms of a POVM $\hat E(t,\gamma)\,dt$, where the possible results
of the measurement are labeled by the continuous parameter $t$ and
the discrete parameter $\gamma=\pm1$.  The POVM satisfies three properties
analogous to Eqs.~(\ref{Extwo}), (\ref{Exone}), and (\ref{Exfour}):
\begin{equation}
\hat E(t,\gamma)\,dt={dt\over C}\,|t,\gamma\rangle\langle t,\gamma|\;,
\label{Etone}
\end{equation}
\begin{equation}
\hat1=
\sum_\gamma\int_{-\infty}^\infty dt\,\hat E(t,\gamma)=
\sum_\gamma\int_{-\infty}^\infty{dt\over C}\,
|t,\gamma\rangle\langle t,\gamma|\;,
\label{Ettwo}
\end{equation}
\begin{equation}
e^{-iT\hat H/\hbar}|t,\gamma\rangle=|t+T,\gamma\rangle\;.
\label{Etfour}
\end{equation}
The displacement condition~(\ref{Etfour}), with $t=0$ and $T=t$, becomes
\begin{equation}
\langle\epsilon,\sigma|t,\gamma\rangle=e^{-it\epsilon/\hbar}
\langle\epsilon,\sigma|t=0,\gamma\rangle\;,
\end{equation}
which leads to
\begin{equation}
\langle\epsilon,\sigma|\!\left(\sum_\gamma\int_{-\infty}^\infty{dt\over C}\,
|t,\gamma\rangle\langle t,\gamma|
\right)\!|\epsilon',\sigma'\rangle=
{2\pi\hbar\over C}\delta(\epsilon-\epsilon')
\sum_\gamma
\langle\epsilon,\sigma|t=0,\gamma\rangle
\langle t=0,\gamma|\epsilon,\sigma'\rangle
\;.
\end{equation}
Thus the completeness condition~(\ref{Ettwo}) can be satisfied by
choosing $C=1$ and by requiring that
\begin{equation}
\sum_\gamma
\langle\epsilon,\sigma|t=0,\gamma\rangle
\langle t=0,\gamma|\epsilon,\sigma'\rangle=
\delta_{\sigma\sigma'}\;,
\end{equation}
which, in turn, means that $\langle\epsilon,\sigma|t=0,\gamma\rangle$
is a $2\times2$ unitary matrix.  By removing the common phase factor from
this unitary matrix, it can be written as
\begin{equation}
\langle\epsilon,\sigma|t=0,\gamma\rangle=
e^{if(\epsilon)}U_{\sigma\gamma}(\epsilon)\;,
\end{equation}
where $U_{\sigma\gamma}(\epsilon)$ is the unit-determinant unitary
matrix of Eq.~(\ref{egamma}).  Notice that the new energy
eigenstates~(\ref{egamma}) satisfy $\langle\epsilon,\gamma|t,\gamma'\rangle=
\delta_{\gamma\gamma'}e^{-it\epsilon/\hbar}$.

Because the energy spectrum is bounded below, the time states
\begin{equation}
|t,\gamma\rangle=
\sum_\sigma\int_0^\infty{d\epsilon\over2\pi\hbar}\,
|\epsilon,\sigma\rangle e^{if(\epsilon)}U_{\sigma\gamma}(\epsilon)
e^{-it\epsilon/\hbar}=
\int_0^\infty{d\epsilon\over2\pi\hbar}\,
|\epsilon,\gamma\rangle e^{-it\epsilon/\hbar}
\label{tstatesgamma}
\end{equation}
are not orthogonal, their inner product being given by
\begin{equation}
\langle t,\gamma|t',\gamma'\rangle=
\delta_{\gamma\gamma'}\int_0^\infty{d\epsilon\over2\pi\hbar}\,
e^{i(t-t')\epsilon/\hbar}=
\delta_{\gamma\gamma'}\Biggl(
{1\over2}\delta(t-t')+{i\over2\pi}{\rm P}\!\left({1\over t-t'}\right)\Biggr)\;.
\end{equation}
The canonical time representation results from choosing $f(\epsilon)=0$
and $U_{\sigma\gamma}(\epsilon)=\delta_{\sigma\gamma}$.

The probability density that a measurement yields results $t$ and $\gamma$,
given parameter $T$, is given by
\begin{equation}
p(t,\gamma|T)=|\psi_T(t,\gamma)|^2=|\psi_0(t-T,\gamma)|^2=p(t-T,\gamma)\;,
\end{equation}
where
\begin{equation}
\psi_T(t,\gamma)=\langle t,\gamma|\psi_T\rangle=
\langle t-T,\gamma|\psi_0\rangle=\psi_0(t-T,\gamma)
\end{equation}
is the wave function of the state $|\psi_T\rangle$ in the time
representation.  The displacement property~(\ref{Etfour}) implies
that in the time representation, $\hat H$ is represented by a derivative:
\begin{equation}
\hat H\Longleftrightarrow{\hbar\over i}{\partial\over\partial t}\;.
\end{equation}

Writing the time wave function of the fiducial state as
$\psi_0(t,\gamma)=r(t,\gamma)e^{i\Theta(t,\gamma)}$, the general
relations~(\ref{meanh}) and (\ref{varianceh}) for the mean and variance
of $\hat h$ become in this case
\begin{equation}
\langle\hat H\rangle=\sum_\gamma\int_{-\infty}^\infty dt\,
\psi_0^\ast(t,\gamma){\hbar\over i}{\partial\over\partial t}\psi_0(t,\gamma)=
\sum_\gamma\int_{-\infty}^\infty dt\,p(t,\gamma)\hbar\Theta'(t,\gamma)\;,
\end{equation}
\begin{eqnarray}
\langle(\Delta\hat H)^2\rangle&=&
\sum_\gamma\int_{-\infty}^\infty dt\,\left|
\left(\hbar{\partial\over\partial t}
-i\langle\hat H\rangle\!\right)\!\psi_0(t,\gamma)\right|^2\nonumber\\
&=&{\hbar^2\over 4}
\sum_\gamma\int_{-\infty}^\infty dt\,{[\,p'(t,\gamma)]^2\over p(t,\gamma)}+
\sum_\gamma\int_{-\infty}^\infty dt\,p(t,\gamma)
[\hbar\Theta'(t,\gamma)-\langle\hat H\rangle]^2\;.\;\;\nonumber\\
&&
\label{varianceH}
\end{eqnarray}
Using this expression for the variance of $\hat H$ or using the general
condition~(\ref{condeq}) for an optimal measurement, one can derive that
the requirement for a global optimal measurement is that
$\Theta(t,\gamma)=\langle\hat H\rangle t/\hbar+\mbox{(constant)}_\gamma$.
Discarding an irrelevant overall phase due to the two constants, but
retaining the differential phase between the $\gamma=\pm1$ parts of the
wave function, one can write the resulting fiducial wave function for a
global optimal measurement as
\begin{equation}
\psi_0(t,\gamma)=e^{i\gamma\mu}r(t,\gamma)e^{i\langle\hat H\rangle t/\hbar}\;,
\label{optwft}
\end{equation}
where $\mu$ is a constant.

The time and energy representations of a state $|\psi\rangle$ are related by
\begin{equation}
\langle t,\gamma|\psi\rangle=\psi(t,\gamma)=
\sum_\sigma\int_0^\infty{d\epsilon\over2\pi\hbar}
e^{it\epsilon/\hbar}e^{-if(\epsilon)}U_{\sigma\gamma}^\ast(\epsilon)
\langle\epsilon,\sigma|\psi\rangle=
\int_0^\infty{d\epsilon\over2\pi\hbar}
e^{it\epsilon/\hbar}\langle\epsilon,\gamma|\psi\rangle\;,
\label{psitE}
\end{equation}
\begin{equation}
\langle\epsilon,\gamma|\psi\rangle=
\sum_\sigma e^{-if(\epsilon)}U_{\sigma\gamma}^\ast(\epsilon)
\langle\epsilon,\sigma|\psi\rangle=
\int_{-\infty}^\infty dt\,e^{-it\epsilon/\hbar}\psi(t,\gamma)\;.
\label{psiEt}
\end{equation}
These representations show that the condition~(\ref{optwft}) for a
global optimal measurement, when written in the energy representation,
with $\epsilon=\langle\hat H\rangle+u$, becomes
\begin{eqnarray}
\Bigl\langle\langle\hat H\rangle+u,\gamma\!\Bigm|\!\psi_0\Bigr\rangle&=&
\sum_\sigma e^{-if(\langle\hat H\rangle+u)}
U_{\sigma\gamma}^\ast\Bigl(\langle\hat H\rangle+u\Bigr)
\Bigl\langle\langle\hat H\rangle+u,\sigma\!\Bigm|\!\psi_0\Bigr\rangle
\nonumber\\
&=&\sum_\sigma e^{if(\langle\hat H\rangle-u)}
U_{\sigma\gamma}\Bigl(\langle\hat H\rangle-u\Bigr)
\Bigl\langle\langle\hat H\rangle-u,\sigma\!\Bigm|\!\psi_0\Bigr\rangle^\ast
\nonumber\\
&=&\Bigl\langle\langle\hat
H\rangle-u,\gamma\!\Bigm|\!\psi_0\Bigr\rangle^\ast\;,
\label{optErep}
\end{eqnarray}
where we discard the differential phase $\mu$ because it can be absorbed
into the unitary matrix $U_{\sigma\gamma}$.  The condition~(\ref{optErep})
can be satisfied, by appropriate choices for the function $f(\epsilon)$
and the unitary matrix $U_{\sigma\gamma}(\epsilon)$, if and only if
\begin{equation}
\sum_\sigma
\Bigl|\Bigl\langle\langle\hat H\rangle+u,\sigma
\!\Bigm|\!\psi_0\Bigr\rangle\Bigr|^2=
\sum_\sigma
\Bigl|\Bigl\langle\langle\hat H\rangle-u,\sigma
\!\Bigm|\!\psi_0\Bigr\rangle\Bigr|^2\;,
\label{optEreptwo}
\end{equation}
i.e., the total probability density to have energy $\langle\hat H\rangle+u$
is the same as the total probability density to have energy
$\langle\hat H\rangle-u$.

\section{Lorentz-Invariant Uncertainty Relations}
\label{lorentzup}

We now apply the formalism developed in Section~\ref{genuppure} to
formulating Lorentz-invariant uncertainty relations for estimation of
the displacement and Lorentz-rotation parameters of the Poincar\'e group.
We deal first with the displacement parameters, where we are seeking
a restriction on the estimation of a space-time translation and, hence,
on the estimation of the invariant space-time interval.  The generator
of space-time translations is the operator for the energy-momentum
4-vector
\begin{equation}
\hat{\bf P}=\hat P^\alpha{\bf e}_\alpha
=\hat P^0{\bf e}_0+\hat{\mbox{$\vec P\mskip 3.5mu$}}\mskip -3.5mu
=\hat P^0{\bf e}_0+\hat P^j{\bf e}_j\;,
\end{equation}
for whatever fields are used to distinguish translated frames.  We write
the displacement 4-vector as
\begin{equation}
{\bf X}=S{\bf n}=Sn^{\alpha}{\bf e}_\alpha \;,
\label{eq21}
\end{equation}
where
\begin{equation}
{\bf n}=n^0{\bf e}_0+\vec n=n^0{\bf e}_0+n^j{\bf e}_j
\end{equation}
is a (timelike or spacelike) unit 4-vector that gives the direction of
the space-time translation and $S$ is the invariant interval that
parametrizes the translation.  The path on Hilbert space is given by
\begin{equation}
|\psi_S\rangle=e^{-iS{\bf n}\cdot\hat{\bf P}/\hbar}|\psi_0\rangle \;,
\end{equation}
where
\begin{equation}
{\bf n}\cdot\hat{\bf P}=\eta_{\alpha\beta}n^\alpha\hat P^\beta
=n^\alpha\hat P_\alpha=-n^0\hat P^0+\vec n\cdot
\hat{\mbox{$\vec P\mskip 3.5mu$}}\mskip -3.5mu
\;.
\end{equation}
Here $||\eta_{\alpha\beta}||={\rm diag}(-1,+1,+1,+1)$ is the Minkowski
metric of special relativity (we adopt units such that the speed of
light $c=1$), and
$\vec n\cdot\hat{\mbox{$\vec P\mskip 3.5mu$}}\mskip -3.5mu=n^j\hat P^j$
is the three-dimensional dot product.

\mbox{}From Eq.~(\ref{pureup}) the uncertainty relation for estimation of
the invariant interval $S$ is
\begin{equation}
\langle(\delta S)^2\rangle_S\langle({\bf n}\cdot\Delta\hat{\bf P})^2\rangle=
\langle(\delta S)^2\rangle_S\,
n^{\alpha}n^{\beta}\langle\Delta\hat P_\alpha\Delta\hat P_\beta\rangle
\ge{\hbar^2\over4N}\;.
\label{intup}
\end{equation}
When ${\bf n}$ is timelike, this is a time-energy uncertainty relation
for the observer whose 4-velocity is ${\bf n}$, and when ${\bf n}$ is
spacelike, this is a position-momentum uncertainty relation for an
observer whose 4-velocity is orthogonal to ${\bf n}$.  In particular,
when ${\bf n}={\bf e}_0$, the time-energy uncertainty relation takes
the form
\begin{equation}
\langle(\delta S)^2\rangle_S\langle(\Delta\hat P^0)^2\rangle
\ge{\hbar^2\over4N}\;,
\end{equation}
and when ${\bf n}=\vec n=n^j{\bf e}_j$ is a spatial unit vector, the
position-momentum uncertainty relation becomes
\begin{equation}
\langle(\delta S)^2\rangle_S\langle(\vec n\cdot
\Delta\hat{\mbox{$\vec P\mskip 3.5mu$}}\mskip -3.5mu)^2\rangle
\ge{\hbar^2\over4N}\;.
\label{relxpup}
\end{equation}

For illustration, suppose that the relevant field is the free
electromagnetic field.  When considering the energy-momentum 4-vector
as a generator, it is most convenient to decompose the field in terms
of plane-wave field modes, for then
\begin{equation}
\hat{\bf P}=\sum_{\vec k,\sigma}\hbar{\bf k}\,
\hat a_{\vec k,\sigma}^\dagger \hat a_{\vec k,\sigma}
\label{4momentum}
\end{equation}
is a sum of separate contributions from the various modes.  In
Eq.~(\ref{4momentum})
${\bf k}=\omega{\bf e}_0+\vec k=\omega{\bf e}_0+k^j{\bf e}_j$ is a
(null) wave 4-vector, with $\omega=|\vec k\,|=k$; the sum runs over
all plane-wave field modes, i.e., over all all wave 3-vectors $\vec k$
and over the two helicities, denoted by $\sigma$.  Since the generator
${\bf n}\cdot\hat{\bf P}/\hbar$ for any space-time translation is
determined by the number operators for the plane-wave field modes,
global optimal measurements will involve measurements of phase shifts
of these modes.  This is not a surprising conclusion because the effect
of a space-time translation is to shift the phase of each plane-wave
field mode.  Indeed, if only a single plane-wave field mode is excited,
the discussion of global optimal measurements of the invariant interval
reduces to the analysis of phase measurement in Section~\ref{phin}.
If many modes are excited, as in a pulse of electromagnetic radiation,
the discussion of global optimal measurements is more complicated.
Measurements of phase shifts in the multi-mode case are only beginning
to be considered \cite{Shapiro,HollandBurnett,SandersMilburn}.  Notice
that when many modes are excited, the generator
${\bf n}\cdot\hat{\bf P}/\hbar$ becomes highly degenerate, a situation
that cannot be addressed by the general considerations of
Section~\ref{optmeas}.

Turn now to the case of Lorentz transformations, where we seek restrictions
on the estimation of the parameters that describe boosts and spatial
rotations.  The generator of Lorentz transformations is the operator for
the antisymmetric angular-momentum two-tensor
\begin{equation}
\hat{\bf J}=\hat J^{\alpha\beta}\hat{\bf e}_\alpha\otimes\hat{\bf e}_\beta\;,
\end{equation}
whose components are given in terms of the stress-energy tensor by
\begin{equation}
\hat{J}^{\alpha\beta}=\int d^3x\left(
x^\alpha\hat T^{\beta0}-x^\beta\hat T^{\alpha0}\right)\;.
\end{equation}
The path on Hilbert space is given by
\begin{equation}
|\psi_\Theta\rangle=\exp\!\left(-{i\over2\hbar}\Theta\,\Omega^{\alpha\beta}
\hat J_{\alpha\beta}\right)|\psi_0\rangle \;,
\end{equation}
where $\Theta$ is the Lorentz-rotation parameter and $\Omega^{\alpha\beta}$
is an antisymmetric two-tensor that specifies the sense of the Lorentz
rotation.

For a boost with velocity $v$ in the direction of a spatial unit vector
$\vec n=n^j{\bf e}_j$, $\Theta$ is the velocity parameter corresponding
to $v$, i.e., $\cosh\Theta=(1-v^2)^{-1/2}$, and the only non-zero components
of $\Omega^{\alpha\beta}$ are the time-space components
$\Omega^{0j}=-\Omega^{j0}=-n^j$.  The path on Hilbert space becomes
\begin{equation}
|\psi_\Theta\rangle=e^{i\Theta\vec n\cdot
{\scriptstyle
\hat{\mbox{${\scriptstyle\vec K\mskip 3.5mu}$}}\mskip -3.5mu}/\hbar}
|\psi_0\rangle\;,
\end{equation}
where the boost generator,
\begin{equation}
\hat{\mbox{$\vec K\mskip 3.5mu$}}\mskip -3.5mu=\hat K^j{\bf e}_j=
\hat J_0\mbox{}^j{\bf e}_j\;,
\end{equation}
is an energy-weighted position operator.  For a spatial rotation
about the spatial unit vector $\vec n=n^j{\bf e}_j$, $\Theta$ is the
rotation angle, and the only non-zero components of
$\Omega^{\alpha\beta}$ are the space-space components
$\Omega^{jk}=\epsilon^{jkl}n^l$, where $\epsilon^{jkl}$ is the
three-dimensional Levi-Civita tensor.  The path on Hilbert space becomes
\begin{equation}
|\psi_\Theta\rangle=e^{-i\Theta\vec n\cdot
{\scriptstyle
\hat{\mbox{${\scriptstyle\vec J\mskip 6.0mu}$}}\mskip -6.0mu}/\hbar}
|\psi_0\rangle\;,
\end{equation}
where the generator of spatial rotations,
\begin{equation}
\hat{\mbox{$\vec J\mskip 6.0mu$}}\mskip -6.0mu=\hat J^j{\bf e}_j=
{1\over2}\epsilon^{jkl}\hat J_{kl}{\bf e}_j\;,
\end{equation}
is the angular-momentum operator.

The general form of the uncertainty relation for estimation of the
Lorentz-rotation parameter $\Theta$ is
\begin{equation}
\langle(\delta\Theta)^2\rangle_\Theta\,
{1\over4}\Omega^{\alpha\beta}\Omega^{\mu\nu}
\langle\Delta\hat J_{\alpha\beta}\Delta\hat J_{\mu\nu}\rangle
\ge{\hbar^2\over4N}\;.
\label{rotup}
\end{equation}
For a boost the uncertainty relation,
\begin{equation}
\langle(\delta\Theta)^2\rangle_\Theta\langle(\vec n\cdot
\Delta\hat{\mbox{$\vec K\mskip 3.5mu$}}\mskip -3.5mu)^2\rangle
\ge{\hbar^2\over4N}\;,
\label{boostup}
\end{equation}
expresses the quantum-mechanical limitations on determining the velocity
parameter $\Theta$.  This uncertainty relation is complementary to
the relativistic position-momentum uncertainty relation~(\ref{relxpup}).
In Eq.~(\ref{relxpup}) the parameter is a spatial displacement, and the
operator is the component of 3-momentum which generates the displacement.
In Eq.~(\ref{boostup}) the parameter is related to a velocity change, and
the operator is the component of energy-weighted position which generates
the change in velocity.  For a spatial rotation the uncertainty relation,
\begin{equation}
\langle(\delta\Theta)^2\rangle_\Theta\langle(\vec n\cdot
\Delta\hat{\mbox{$\vec J\mskip 6.0mu$}}\mskip -4.0mu)^2\rangle
\ge{\hbar^2\over4N}\;,
\end{equation}
expresses the quantum-mechanical limitations on determining a rotation.

To investigate global optimal measurements of a spatial rotation or a
boost, it would be wise to decompose the relevant field in terms of
angular-momentum modes or ``boost modes.''  Such an investigation lies
outside the scope of the present paper.

\section{Conclusion}
\label{conclusion}

Much ink has been devoted to the problem that many quantities of physical
interest, such as time or harmonic-oscillator phase, though determined
routinely from measurements, cannot be accommodated within the conventional
quantum-mechanical description of measurements, because such quantities
have no associated Hermitian operator.  The aim of this paper is to show
that this problem is only apparent.  We eschew tedious discussions of the
status of such quantities in quantum theory.  Instead we develop a
formalism that allows us to derive quantum-mechanical limitations on
the determination of such a quantity, without ever having to introduce
an operator associated with the quantity, and we illustrate the formalism
with numerous examples.

The formalism is founded on the idea that such a quantity should be
treated as a parameter, to be determined from the results of measurements.
To derive strict quantum-mechanical limits on such a determination, we
must be able, first, to describe all measurements permitted by the rules
of quantum mechanics---this is accomplished by using the formalism of
POVMs---and, second, to set bounds on all possible ways of estimating the
parameter from the results of the measurements---this is accomplished
by appealing to the Cram\'er-Rao bound of classical parameter-estimation
theory.  The resulting quantum-mechanical limitations are expressed
as Mandelstam-Tamm uncertainty relations involving the precision of the
parameter estimation and variance of the operator that generates changes
in the parameter.  These uncertainty relations take into account naturally
the expected improvement in determining the parameter as one is allowed
to make measurements on an increasing number of identically prepared
systems.  Moreover, we are able to derive general conditions for optimal
measurements that can achieve the lower bound in the uncertainty relation,
although it is generally not known how to perform such optimal measurements.
The final result is a formalism that increases considerably the scope
and power of uncertainty relations in quantum theory.

\newpage

\begin{figure}
\caption{ Phase-plane representation of optimal measurements of
displacement of a squeezed vacuum state.  The squeezed vacuum state
$|\psi_0\rangle=\hat S|{\rm vac}\rangle$ is represented by a solid
``uncertainty ellipse'' centered at the origin.  The principal axes
of the ellipse are oriented along the directions defined by the
uncorrelated co\"ordinates $\protect\underline{x}'$ and $p'$, which
are rotated by an angle $\varphi$ relative to the axes defined by the
canonical position $\protect\underline{x}$ and the momentum $p$; the
principal radii of the ellipse are given by the uncertainties
$\langle\psi_0|(\Delta\protect\underline{\hat x}')^2|\psi_0\rangle^{1/2}=
e^{-r}/\protect\sqrt{2}$ and
$\langle\psi_0|(\Delta\hat p')^2|\psi_0\rangle^{1/2}/\hbar=
e^r/\protect\sqrt{2}$.
The dotted uncertainty ellipse depicts the state obtained by displacing
the squeezed vacuum state a distance $X$ along the $\protect\underline{x}$
axis.  The global optimal measurement for distinguishing displaced
squeezed states corresponds to measuring a variable $x$ [see
Eq.~(\protect\ref{optop2})], which is a rescaled position variable
along an axis rotated by angle $\theta$ from the axis of the canonical
position variable $\protect\underline{x}$.  The optimal measurement
represents a compromise between maximal ``signal,'' which would be obtained
by measuring the canonical position variable $\protect\underline{x}$, and
minimal ``noise,'' which would be obtained by measuring the rotated
position variable $\protect\underline{x}'$ [see Eq.~(\protect\ref{minnsr})].}
\label{onlyfig}
\end{figure}

\end{document}